\documentclass[preprint]{elsarticle}
\usepackage{xcolor}
\usepackage{lineno,hyperref}
\usepackage[margin=1.2in]{geometry}
\usepackage{amsmath, amsbsy, amsthm, array, mathrsfs}
\usepackage{graphics}
\usepackage{physics}
\usepackage{epsfig, amssymb,latexsym,verbatim}
\usepackage{graphicx}
\usepackage{color}
\usepackage{dsfont, bbm}
\usepackage{relsize}
\usepackage{subcaption}

\newcommand{\R}{\mathbb{R}}
\newcommand{\mbs}{\mathbb{S}}

\newcommand{\noin}{\noindent}
\newcommand{\bee}{\begin{eqnarray*}}
\newcommand{\ene}{\end{eqnarray*}}
\newcommand{\bec}{\begin{center}}
\newcommand{\enc}{\end{center}}
\newcommand{\be}{\begin{equation}}
\newcommand{\ee}{\end{equation}}

\newcommand{\mc}{\mathcal}

\newcommand{\ep}{\varepsilon}
\newcommand{\mb}{\mathbf}
\newcommand{\bs}{\boldsymbol}
\newcommand{\tb}{\textbf}
\newcommand{\pend}{$\blacksquare$}
\newcommand{\vs}{\vskip 3mm}
\newcommand{\bi}{\begin{itemize}}
\newcommand{\ei}{\end{itemize}}
\begin{document}

\begin{frontmatter}
\title{\LARGE  Computation of projection regression depth and its induced median
 \\[4ex]
}

\author{Yijun Zuo\corref{mycorrespondingauthor}} 
\cortext[mycorrespondingauthor]{Corresponding author}
\address{Department of Statistics and Probability,  Michigan State University, East Lansing, MI 48824, USA} 
         \ead{zuo@msu.edu} 

\begin{abstract}
Notions of depth in regression have been introduced and studied in the literature.
The most famous example is Regression Depth (RD), which is a
direct extension of location depth to regression. The projection regression depth (PRD) is the extension of another prevailing location depth, the projection depth, to regression.
 The computation issues of the RD
 have been discussed in  the literature. 
 The computation issues of the PRD have never been dealt with before.
  The computation issues of the PRD and its induced median (maximum depth estimator) in a regression setting are addressed now. 
   For a given $\bs{\beta}\in\R^p$ exact algorithms for the PRD with cost $O(n^2\log n)$ ($p=2$) and $O(N(n, p)(p^{3}+n\log n+np^{1.5}+npN_{Iter}))$ ($p>2$) and approximate algorithms for the PRD and its induced median with cost respectively $O(N_{\mb{v}}np)$ and $O(Rp N_{\bs{\beta}}(p^2+nN_{\mb{v}}N_{Iter}))$ are proposed. Here $N(n, p)$ is a number defined based on the total number of $(p-1)$ dimensional hyperplanes formed by points induced from sample points and the $\bs{\beta}$;  $N_{\mb{v}}$ is the total number of unit directions $\mb{v}$ utilized; $N_{\bs{\beta}}$ is the total number of candidate regression parameters $\bs{\beta}$ employed; $N_{Iter}$ is the total number of iterations carried out in an optimization algorithm; $R$ is the total number of replications.
Furthermore, as the second major contribution, three PRD induced estimators, which can be computed up to 30 times faster than that of the PRD induced median while maintaining a similar level of accuracy are introduced.
Examples and simulation studies reveal that the depth median induced from the PRD is favorable in terms of robustness and efficiency, compared to the maximum depth estimator induced from the RD, which is the current leading regression median.
\end{abstract}

\end{frontmatter}

\section{Introduction}
Notions of location depth have been introduced and extensively studied in the literature over the last three decades. Depth notions have found applications in diverse fields and disciplines (see Zuo (2018a) for a review).
Among others (Simplicial depth (Liu (1990)), Zonoid depth (Koshevoy and Mosler (1997), Mosler (2002, 2012)) and Spatial depth (Vardi and Zhang (2000), etc.),  two prevailing location depth notions are the Tukey halfspace depth (HD) (Tukey (1975)) (popularized by Donoho and Gasko (1992)) and the projection depth (PD) (Liu(1992), Zuo and Serfling (2000)) (thoroughly studied in Zuo (2003)), both of which are in the spirit of the projection-pursuit scheme.
\vs
One naturally wonders if the depth notion can be extended to a regression setting. Regression depth (RD) of Rousseeuw and Hubert (1999) (RH99) is the most famous example, which directly extends HD to regression, whereas projection regression depth (PRD),  induced from Marrona and Yohai (1993) (MY93) and introduced in Zuo (2018a) (Z18a), is an extension of the PD to regression.
\vs
 Like their location counterparts, the most remarkable advantage of the notion of depth in regression is the direct introduction of the median-type estimator, otherwise known as the maximum (or deepest) regression depth estimator for regression parameters in a multi-dimensional setting. The maximum (deepest) regression depth estimators serve as \emph{robust} alternatives to the classical least squares or least absolute deviations estimator of  unknown parameters in a general linear regression model:\vspace*{-2mm}
\begin{eqnarray}
y&=& (1, \mathbf{x}')\boldsymbol{\beta}+{{e}}, \label{eqn.model}
\end{eqnarray}
where  $'$ denotes the transpose of a vector, and random vector $\mathbf{x}=(x_1,\cdots, x_{p-1})'\in \R^{p-1}$ and  parameter vector $\boldsymbol{\beta}=(\beta_0,\bs{\beta}'_1)' \in \R^p$ ($p\geq2$) and random variables $y$ and ${e}$ are in $\R^1$.
Let $\mb{w}=(1,\mb{x}')'$. Then $y=\mb{w}'\bs{\beta}+{e}$. We use this model or (\ref{eqn.model}) interchangeably depending on the context.
\vs
The maximum depth estimator induced from the RD, $\mb{T}^*_{RD}$, can asymptotically resist up to $33\%$ (Van Aelst and Rousseeuw (2000) (VAR00))(whereas the one from the PRD, $\mb{T}^*_{PRD}$, can resist up to $50\%$ (Zuo (2019a)(Z19a)) contamination without breakdown,
in contrast to the $0\%$ of the classical LS estimator. An illustration of these facts is given in Figure 1, where the data set is given in Table 9 of Chapter 2 from Rousseeuw and Leroy (1987) (RL87). The original data set contains nine bivariate points.
\bec
\begin{figure}[ht]
    \centering
    \begin{subfigure}[ht]{0.3\textwidth}
        \includegraphics[width=4cm, height=4cm]{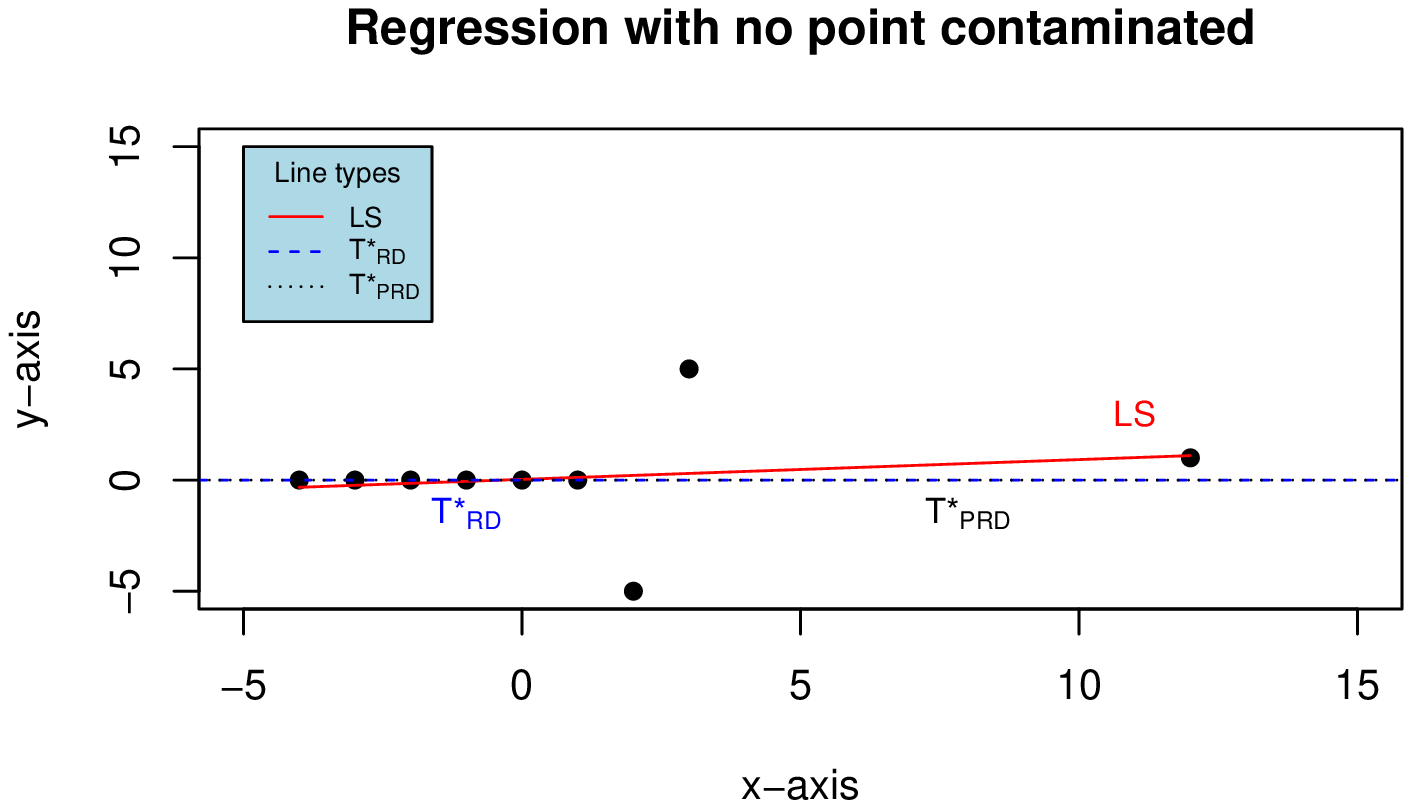}
        \caption{no-contamination}
        \label{fig:no-camtami}
    \end{subfigure}
     \begin{subfigure}[ht]{0.3\textwidth}
        \includegraphics[width=4cm, height=4cm]{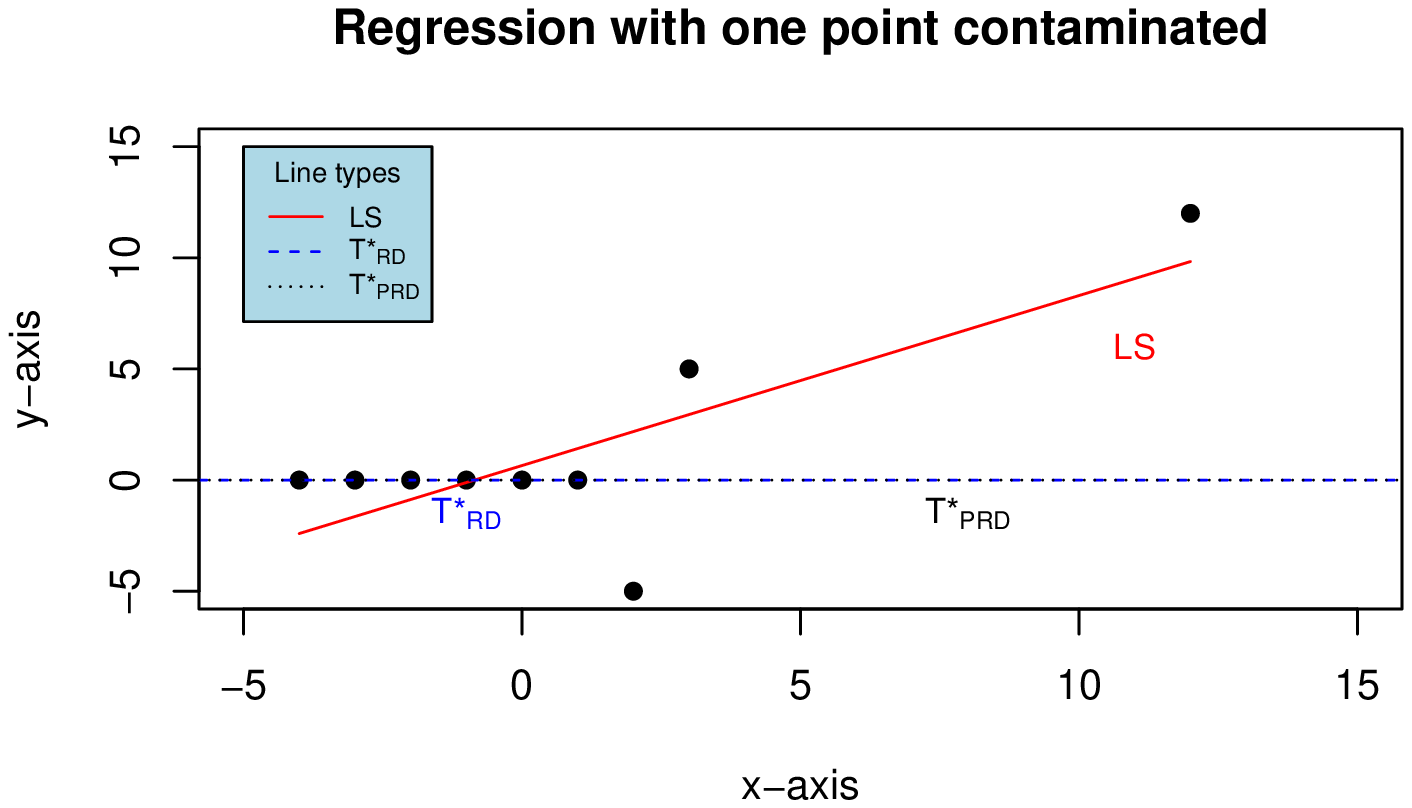}
        \caption{1-point-contaminated}
        \label{fig:one-contami}
    \end{subfigure}
    \begin{subfigure}[ht]{0.3\textwidth}
        \includegraphics[width=4cm, height=4cm]{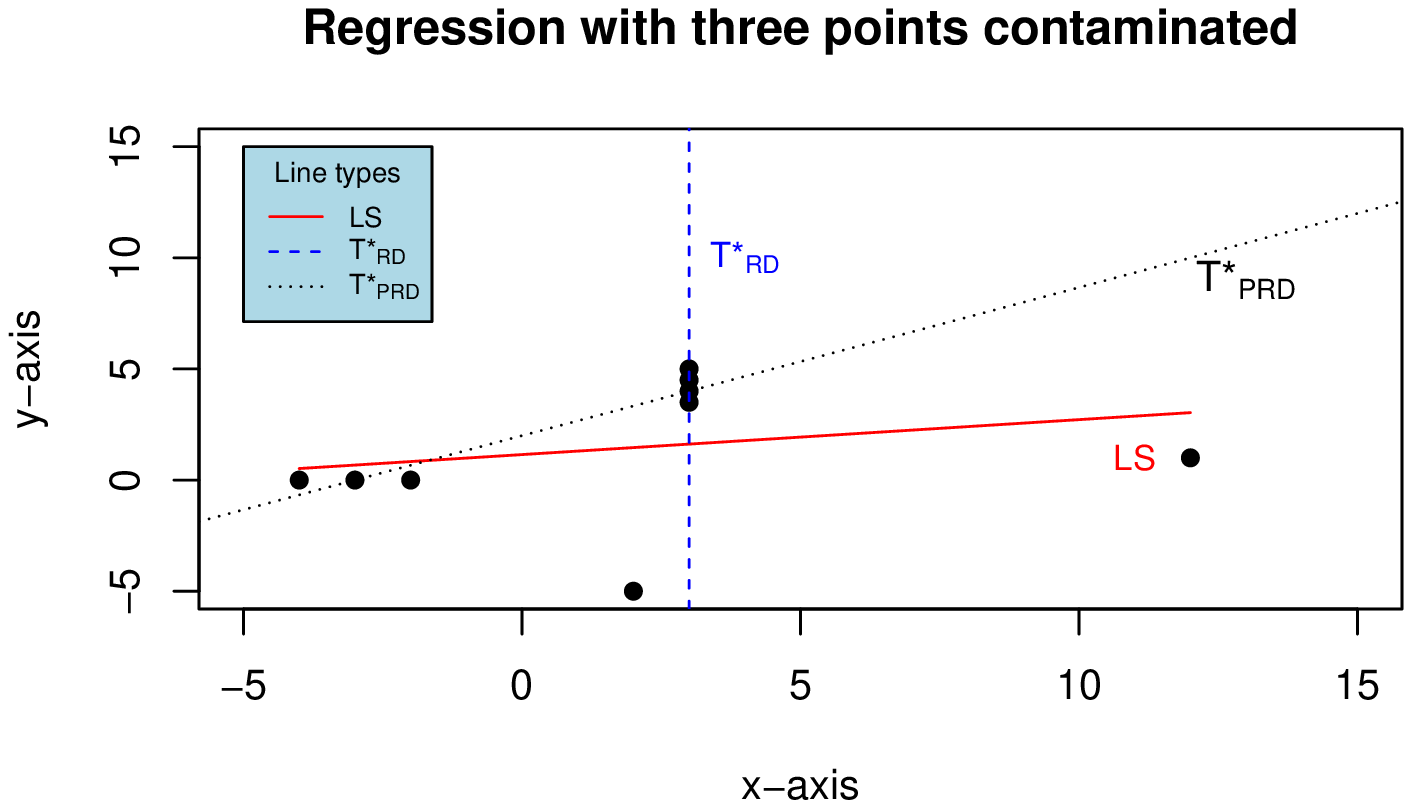}
        \caption{3-points-contamination}
        \label{fig:contami}
    \end{subfigure}
    \caption{
    Three regression lines for data with or without contamination (solid red for the LS, dashed blue for the $\mb{T}^*_{RD}$ and dotted black for the $\mb{T}^*_{PRD}$). (a) Original nine-point data set, the $\mb{T}^*_{RD}$ and the $\mb{T}^*_{PRD}$ are identical. (b) Contaminated data set with one original point $(12,1)'$ moved to $(12,12)'$, leading to a drastic change in the LS line while both the $\mb{T}^*_{RD}$ and the $\mb{T}^*_{PRD}$ are unchanged and resist the contamination.
     (c) Contaminated data set with three original points moved to the points with 3 as their x-coordinates, the $\mb{T}^*_{RD}$ breaks down while both the $\mb{T}^*_{PRD}$ and the LS lines are still informative.}
\end{figure}
\vspace*{-10mm}
\enc
\vs
~~~~~For any $\bs{\beta}\in \R^p$ and joint distribution $P$ of $(\mb{x}', y)$ in $\R^{p}$, RH99 defined the \emph{regression depth} of
 $\bs{\beta}$--denoted hereafter by RD$(\bs{\beta};P)$--
to be the minimum probability mass that needs to be passed when tilting (the hyperplane induced from) $\bs{\beta}$ in any way until it is vertical. The maximum regression depth estimating functional $\mb{T}^*_{RD}$ (also denoted by $\bs{\beta}^*_{RD}$) is then defined as
\be \mb{T}^*_{RD}(P)=\arg\!\max_{\bs{\beta}\in\R^p}RD(\bs{\beta};P). \label{T-RD.eqn}
\ee
Various characterizations of RD$(\bs{\beta};P)$ have been given in the literature, e.g. Zuo (2018b).
\vs
By modifying the P-estimate of Marrona and Yohai (1993) (MY93) to achieve the scale invariance property, Z18a introduced  projection regression depth (PRD), defined based on the so-called ``unfitness" (UF) for a given candidate regression parameter $\bs{\beta}\in \R^p$:
\be
\mbox{UF}(\bs{\beta};F_{(\mb{x}', y)})=\sup_{\mb{v}\in \mbs^{p-1}}|{R}(F_{(\mb{w'}\mb{v},~ y-\mb{w'}\bs{\beta})})|\big/{S}(F_y),  \label{UF.eqn}
\ee
\be
\mbox{PRD}(\bs{\beta};F_{(\mb{x}', y)})=1/(1+\mbox{UF}(\bs{\beta};F_{(\mb{x}', y)})),\label{PRD.eqn}
\ee
\vs
\noin
where $F_{\mb{Z}'}$ stands for the distribution of the p-dimensional random vector ${\mb{Z}} \in\R^p$,
$\mb{w}=(1,\mb{x}')'\in\R^p$, $\mbs^{p-1}=\{\mb{u}\in\R^p:~\|\mb{u}\|=1\}$, $R$ will be restricted to the univariate regression functional  of the form $R(F_{(\mb{w'}\mb{v},~ y-\mb{w'}\bs{\beta})})=T\big(F_{\frac{y-\mb{w}'\bs{\beta}}{\mb{w}'\mb{v}}}\big)$ and it is regression, scale, and affine equivariant (see page 116 of RL87 for definitions). $T$ could be a univariate location functional that is location, and scale (or called affine) equivariant; $S$ is a scale functional that is translation invariant and scale equivariant (see pages 158-159 of RL87 for definitions), and $S(F_y)$ does not depend on $\mb{v}$ and $\bs{\beta}$, see Z18a.
\vs
It is not difficult to see that $\mbox{UF}(\bs{\beta}; F_{(\mb{x}', y)})$ and $\mbox{PRD}(\bs{\beta};F_{(\mb{x}', y)})$ are the regression counterparts of the outlyingness function $O(\bs{\beta};F_{\mb{x}})$ and the projection depth function $\mbox{PD}(\bs{\beta}; F_{\mb{x}})$ (Zuo (2003)), respectively.
\vs
Examples of $T$ in (\ref{UF.eqn}) include mean, quantile, median (Med),  and location functionals in Wu and Zuo (2009). Examples of $S$ in (\ref{UF.eqn}) include standard deviation,  median absolute deviations from the median (MAD), and scale functionals in Wu and Zuo (2008).
\vs
  For the consideration of robustness, in the sequel, $(T,S)$ is fixed and it is
the pair $(\mbox{Med}, \mbox{MAD})$, 
unless otherwise stated. Hereafter, we write $\text{Med}(Z)$ rather than
  $\text{Med}(F_Z)$.
For this special choice of $T$  and $S$  such that
\bee
 R(F_{(\mb{w'}\mb{v},~ y-\mb{w'}\bs{\beta})})&=&\text{Med}_{\mb{w'}\mb{v}\neq 0}\big(\frac{y-\mb{w}'\bs{\beta}}{\mb{w'}\mb{v}}\big),\\[1ex]\label{spesific-T.eqn}
 S(F_y)&=& \text{MAD}(F_y). \label{S.eqn}
 \ene
We have
 \be
 \text{UF}(\bs{\beta}; F_{(\mb{x}', y)})=\sup_{\mb{v}\in \mbs^{p-1}}\Big|\text{Med}_{\mb{w'}\mb{v}\neq 0}\big(\frac{y-\mb{w}'\bs{\beta}}{\mb{w'}\mb{v}}\big)\Big|\bigg/ \text{MAD}(F_y),
 \ee
 and
 \be
 \text{PRD}\left(\bs{\beta}; F_{(\mb{x}', y)}\right)=\inf_{\mb{v}\in \mbs^{p-1},\mb{w'}\mb{v}\neq 0}
 \frac{\text{MAD}(F_y)}{\text{MAD}(F_y)+\Big|\text{Med}\big(\frac{y-\mb{w}'\bs{\beta}}{\mb{w'}\mb{v}}\big)\Big|}. \label{special-PRD.eqn}
 \ee
 \vs
 \noin
 Applying the min-max (or max-min) scheme, we obtain the maximum (deepest) \emph{projection regression depth estimating functional} (also denoted by $\bs{\beta}^*_{\text{PRD}}$) w.r.t. the pair $(T,S)$
 \begin{eqnarray}
 \mb{T}^*_{PRD}(F_{(\mb{x}', y)})&=&\arg\!\min_{\boldsymbol{\beta}\in \R^p}\mbox{UF}(\boldsymbol{\beta}; ~F_{(\mb{x}', y)})  \label{eqn.T*} 
 =\arg\!\max_{\bs{\beta}\in\R^p}\text{PRD}\left(\boldsymbol{\beta}; ~F_{(\mb{x}', y)}\right). \label{eqn.T*} 
 \end{eqnarray}
\vs
When a sample ${\mb{Z}}^n=\{(\mb{x}'_i, y_i)', i=1,\cdots, n\}$ of ${\mb{Z}}:=(\mb{x}',y)' \in \R^{p}$ is given, an empirical distribution $F^n_{\mb{Z}}$ based on ${\mb{Z}}^n$ is obtained. Replacing $F_{(\mb{x}',y)}$ above by $F^n_{\mb{Z}}$ we obtain all empirical versions. 
\vs

While both the RD and the PRD enjoy desirable properties such as high breakdown robustness, these regression depth functions prove difficult to compute in practice since they involve the projection-pursuit scheme (see Z18a). The computation of the RD has been discussed in RH99, in  Rousseeuw and Struyf (1998) (RS98), and in Liu and Zuo (2014) (LZ14). The computation issues of the PRD and the $\mb{T}^*_{PRD}$ have never been addressed. Presenting exact and approximate algorithms for the PRD and discussing the algorithms for the computation of the $\mb{T}^*_{PRD}$ are the two main goals of this article. The third goal is to introduce several PRD induced estimators which can be computed much faster than that of the $\mb{T}^*_{PRD}$.
\vs

The rest of this article is organized as follows. Section 2 presents the computation problem and addresses the exact and approximate computation algorithms for the UF$(\bs{\beta}, F^n_{\mb{Z}})$, and equivalently for the PRD$(\bs{\beta}, F^n_{\mb{Z}})$. Furthermore,
theoretical results are established and exact and approximate algorithms are presented along with abounded examples. 
Section 3 is devoted to (i) the computation of the $\mb{T}^*_{PRD}(\bs{\beta}, F^n_{\mb{Z}})$,
(ii) examples of the exact computation of the PRD as well as the approximate computation of the $\mb{T}^*_{PRD}$, and (iii) comparisons of performance between the $\mb{T}^*_{PRD}$ against
leading competitors such as LS, $\bs{\beta}^*_{RD}$ , and ltsReg.
Section 4 introduces three depth induced regression estimators that can run much faster than $\mb{T}^*_{PRD}$  in addition to maintaining  small empirical mean squared errors. Section 5 investigates the finite sample relative efficiency of the $\mb{T}^*_{PRD}$.
 Brief concluding comments end Section 6 and the article.
\section{Computation of PRD}
\vs

\subsection{The computation problem}

To compute the $\mbox{PRD}(\bs{\beta}; F^n_{\mb{Z}})$, it suffices to compute the $\mbox{UF}(\bs{\beta}; F^n_{\mb{Z}})$. 
Namely, to compute the following quantity:
\be
\mbox{UF}(\bs{\beta}; F^n_{\mb{Z}})=\sup_{\mb{v}\in \mbs^{p-1}}\Big|\text{Med}_{\mb{w}'_i\mb{v}\neq 0}\big\{\frac{y_i-\mb{w}'_i\bs{\beta}}{\mb{w}'_i\mb{v}}\big\}\Big|\bigg/S_y, \label{UF-1.eqn}
\ee
where $\mb{w}'_i=(1,\mb{x}'_i)$ and $S_y=\text{MAD}\{y_i\}$.  Hereafter, we assume that \tb{(A1)}: $P(\mb{w}'\mb{v}=0)=0$,~$\forall~ \mb{v}\in\mbs^{p-1}$ 
 and \tb{(A2)} $P(r(\bs{\beta})=0)=0$, where $r(\bs{\beta})=y-\mb{w}'\bs{\beta}$,
~$\forall ~\bs{\beta}\in \R^p$.
\tb{(A1)-(A2)} hold automatically if $(\mb{x}', y)'$ has a density or if $\mb{x}$ does not concentrate on a single $(p-2)$ dimensional hyperplane in $\mb{x}$ space and any $(p-1)$ dimensional hyperplane determined by $r(\bs{\beta})=0$ in $(\mb{x}', y)'$ space does not contain any probability mass.
\vs

For simplicity of description, we write $\mb{t}'_i=\mb{w}_i'/r_i(\bs{\beta})$,  where 
$r_i(\bs{\beta})=y_i-\mb{w}'_i\bs{\beta}$. Now the computation of $\mbox{UF}(\bs{\beta}; F^n_{\mb{Z}})$ in (\ref{UF-1.eqn}) is equivalent to
the computation of
\be
\mbox{UF}(\bs{\beta}; F^n_{\mb{Z}})=\sup_{\mb{v}\in \mbs^{p-1}}\bigg|\text{Med}_{\mb{t}'_i\mb{v}\neq 0 }\big\{\frac{1}{\mb{t}'_i\mb{v}}\big\}\bigg|\bigg/S_y .\label{UF-2.eqn}
\ee

Again for simplicity, we write $k^{\mb{v}}_i=1/\mb{t}_i'\mb{v}$ and $u^{\mb{v}}_i=\mb{t}_i'\mb{v}$. The latter two are well defined almost surely (a.s.) under \tb{(A1)-(A2)}.  Without loss of generality (w.l.o.g.), 
we assumes  that $S_y=1$ (since it does not depend on $\mb{v}$ or $\bs{\beta}$). The UF$(\bs{\beta};F^n_{\mb{Z}})$ in (\ref{UF-2.eqn}) is then 
\be
\mbox{UF}(\bs{\beta}; F^n_{\mb{Z}})=\sup_{\mb{v}\in\mbs^{p-1}}\bigg|\text{Med}_i\big\{k^{\mb{v}}_i\big\} \bigg| :=\sup_{\mb{v}\in\mbs^{p-1}}\big|g(\mb{v}) \big|.
\label{UF-3.eqn}
\ee
\vs
The exact computation of  (\ref{UF-3.eqn}) above is still very challenging if not impossible.
Let $k^{\mb{v}}_{(1)}\leq k^{\mb{v}}_{(2)}\cdots\leq k^{\mb{v}}_{(n)}$ be ordered values of $k^{\mb{v}}_i$.
 Partition $\mbs^{p-1}$ into two disjoint parts
\be
\mathscr{S}_1=\{\mb{v}\in \mbs^{p-1}:~ k^{\mb{v}}_{(1)}<0 ~\mbox{and}~ k^{\mb{v}}_{(n)}>0\};~~~\mathscr{S}_2=\{\mb{v}\in\mbs^{p-1}:~ k^{\mb{v}}_{(1)}> 0 ~\mbox{or}~ k^{\mb{v}}_{(n)}< 0\}. \label{Si.eqn}
\ee
It is readily seen that both $\mc{S}_1$ and $\mc{S}_2$ are symmetric about the origin. That is, if $\mb{v} \in \mc{S}_i$ then, $\mb{-v} \in \mc{S}_i$.
Now the UF$(\bs{\beta};F^n_{\mb{Z}})$ in (\ref{UF-3.eqn})  can be expressed as follows:
\be
\mbox{UF}(\bs{\beta}; F^n_{\mb{Z}})=\max\big\{\sup_{\mb{v}\in \mc{S}_1}|g(\mb{v})|, ~\sup_{\mb{v}\in \mc{S}_2}|g(\mb{v})|\big\}. \label{UF-4.eqn}
\ee

\vs
Exact computation of the UF$(\bs{\beta};F^n_{\mb{Z}})$ in (\ref{UF-4.eqn}) is a challenging task, whereas approximate computation is relatively straightforward (but it is still difficult to assess its accuracy without the benchmark of the exact result). We shall address the two approaches separately in the sequel.

\subsection{Exact Computation}

\subsubsection{Theoretical results}
\vs
\noin
For a given sample $\mb{Z}^{(n)}:=\{(\mb{x}'_i, y_i)', ~ i=1,\cdots, n\}$, a $\bs{\beta}$ in $\R^p$ and a $\mb{v}\in\mbs^{p-1}$, since $k^{\mb{v}}_{(1)}\leq k^{\mb{v}}_{(2)}\leq \cdots \leq k^{\mb{v}}_{(n)}$ are ordered values of $k^{\mb{v}}_{i}= 1/\mb{t}'_i\mb{v}$, then $1/\mb{t}'_{i_1}\mb{v}\leq 1/\mb{t}'_{i_2}\mb{v}\leq \cdots \leq 1/\mb{t}'_{i_n}\mb{v}$ for some $\{i_1,\cdots, i_n\}$, a permutation of $\{1, 2, \cdots, n\}$.
 Similarly, $u^{\mb{v}}_{(1)}\leq u^{\mb{v}}_{(2)}\leq \cdots \leq u^{\mb{v}}_{(n)}$  corresponds to a permutation $\{j_i,\cdots, j_n\}$ such that
  $u^{\mb{v}}_{j_1}\leq u^{\mb{v}}_{j_2}\leq \cdots,\leq u^{\mb{v}}_{j_n}$ for  $u^{\mb{v}}_i=\mb{t}'_i\mb{v}$.

 \vs
\noin
\tb{Proposition 2.1}: Assume \tb{(A1)} and \tb{(A2)} hold. Let $N^{-}_{\mb{v}}:=\sum_{i=1}^n\mb{I}(k_i^{\mb{v}}<0)$.  The unfitness of $\bs{\beta}$ in (\ref{UF-1.eqn}) can be computed equivalently via 
(\ref{UF-4.eqn}) which 
 can be computed
as follows.
\vs
\noin
Denote $n1:=\lfloor(n+1)/2\rfloor$ and $n2:=\lfloor(n+2)/2\rfloor$, where $\lfloor \cdot\rfloor$ is the floor function.  \vs
\noin
(i) For $\mb{v} \in \mc{S}_2$, 
\[
\sup_{\mb{v}\in \mc{S}_2}|g(\mb{v})|= \left\{
\begin{array}{ll}
\max_{\mb{v}\in \mc{S}_2} \frac{(\mb{t}'_{i_{n1}}+\mb{t}'_{i_{n2}})\mb{v}\big/2} {\mb{v}'\mb{t}_{i_{n1}}\mb{t}'_{i_{n2}}\mb{v}}
 &\mbox{~if~} N^{-}_{\mb{v}}=0, 
\\[3ex]
-\min_{\mb{v}\in \mc{S}_2} \frac{(\mb{t}'_{i_{n1}}+\mb{t}'_{i_{n2}})\mb{v}\big/2} {\mb{v}'\mb{t}_{i_{n1}}\mb{t}'_{i_{n2}}\mb{v}} &\mbox{~if~} N^{-}_{\mb{v}}=n. 
\end{array}
\right.
\label{UF-34.eqn}
\]
\vs
\noin
(ii) For $\mb{v}\in \mc{S}_1$, let $m$ be a non-negative integer.
\begin{itemize}
\item[]
 if $n=2m+1$,
\[
\sup_{\mb{v}\in \mc{S}_1}|g(\mb{v})|=\left\{
\begin{array}{ll}
-1\Big/\max_{\mb{v}\in \mc{S}_1} \mb{t}'_{i_{n1}}\mb{v}      
& \mbox{if $k^{\mb{v}}_{(n1)}<0$},\\[2ex]
1\Big/\min_{\mb{v}\in \mc{S}_1}\mb{t}'_{i_{n1}}\mb{v}                     
& \mbox{if $k^{\mb{v}}_{(n1)}>0$},
\end{array}
\right.
\]
\item[] if $n=2m+2$,
\[
\sup_{\mb{v}\in \mc{S}_1}|g(\mb{v})|=\left\{
\begin{array}{ll}
\bigg|\max_{\mb{v}\in \mc{S}_1} \frac{(\mb{t}'_{i_{n1}}+\mb{t}'_{i_{n2}})\mb{v}\big/2}{ \mb{v}'\mb{t}_{i_{n1}}\mb{t}'_{i_{n2}}\mb{v} }  \bigg| &
~\mbox{if}~~ k^{\mb{v}}_{(n1)}<0 ~\mbox{and}~ k^{\mb{v}}_{(n2)}>0, \\[2.5ex]
\max_{\mb{v} \in \mc{S}_1} \frac{\big(\mb{t}'_{i_{n1}}+\mb{t}'_{i_{n2}}\big)\mb{v}\big/2}
{\mb{v}'\mb{t}_{i_{n1}}\mb{t}'_{i_{n2}} \mb{v}}
 & ~\mbox{if}~~k^{\mb{v}}_{(n1)}>0,\\[3.5ex]
-\min_{\mb{v}\in \mc{S}_1}
\frac{\big(\mb{t}'_{i_{n1}}+\mb{t}'_{i_{n2}}\big)\mb{v}\big/2}
{\mb{v}'\mb{t}_{i_{n1}}\mb{t}'_{i_{n2}} \mb{v}}
& ~\mbox{if}~~ k^{\mb{v}}_{(n2)}<0.
\end{array}
\right.
\]
\end{itemize}
\vs

\noindent
\tb{Proof:}  Note that  under \tb{(A1)-(A2)} $\mc{S}_i$ ($i=1,2$) are a. s. closed sets.  In light of the definitions of the regular univariate sample median and ${\mc{S}}_i$,
the proof follows immediately.
Details are straightforward to verify and thus are omitted. \hfill \pend

\vs
\noindent
\vs
\begin{figure}[t]
    \centering
    \begin{subfigure}[t]{0.47\textwidth}
        \includegraphics[width=\textwidth]{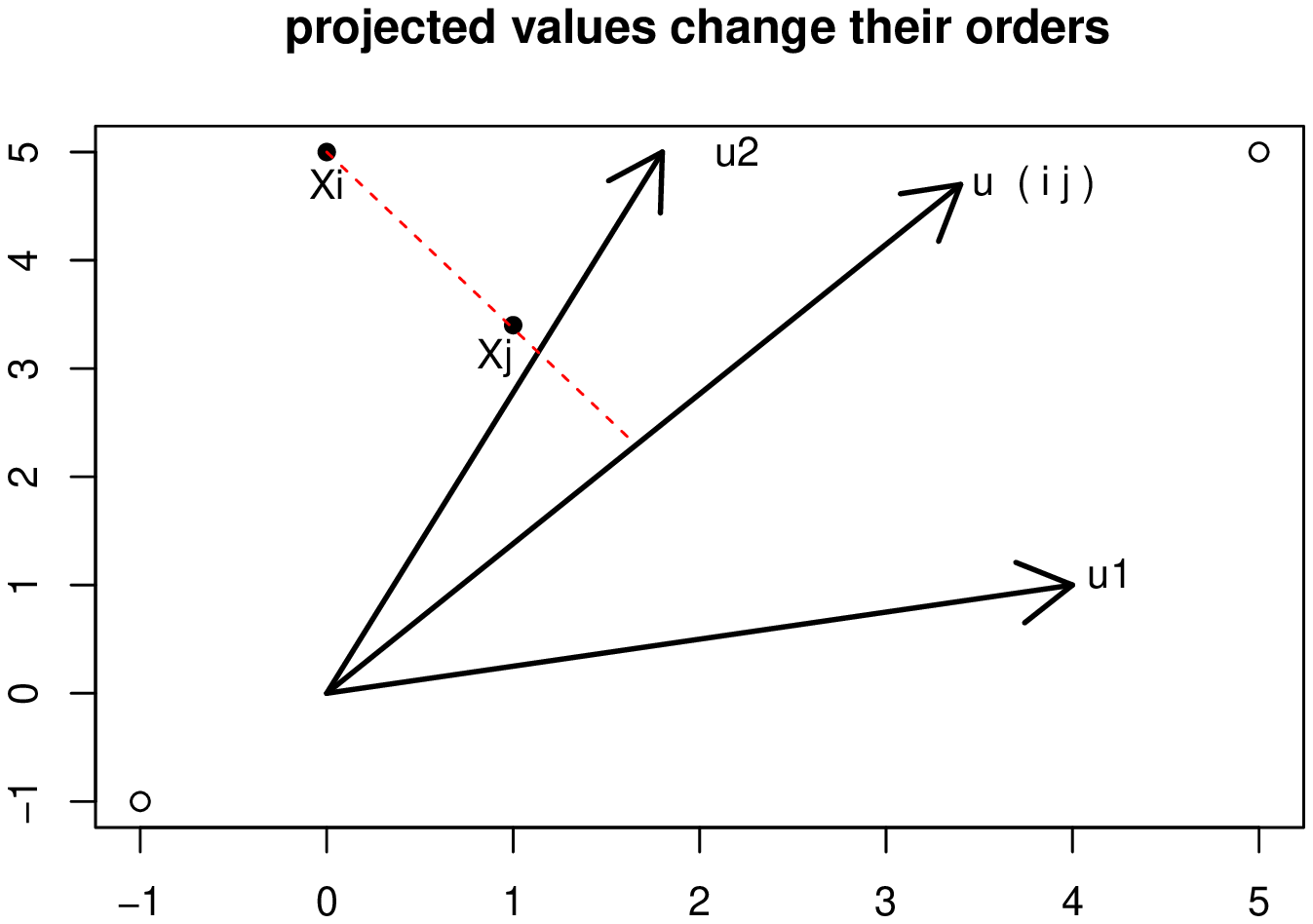}
        \caption{circular-sequence demonstration}
        \label{fig:circular-sequence}
    \end{subfigure}
    \begin{subfigure}[t]{0.47\textwidth}
        \includegraphics[width=\textwidth]{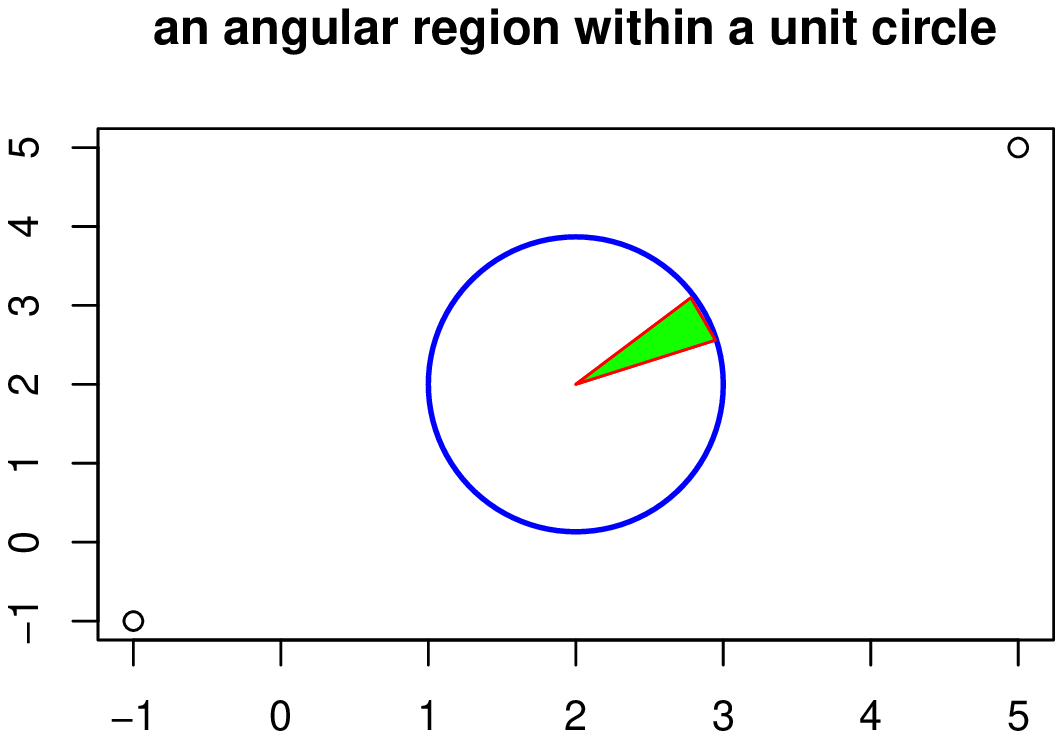}
        \caption{angular region demonstration}
        \label{fig:angular-region}
    \end{subfigure}
    \caption{ {
     (a) $\mb{u}$ (or ij) is perpendicular to the line segment connecting the points $X_i$, $X_j$ and between  $\mb{u}_1$ and $\mb{u}_2$. When the two points are projected to $\mb{u}_1$, $\mb{u}_2$ and $\mb{u}$, the $X_i$ precedes the $X_j$ on $\mb{u}_1$ whereas on $\mb{u}_2$ it is reversed. On $\mb{u}$ they overlap.  
     (b) a unit circle is cut into pieces (angular regions) by the median sequence. Over each piece, the median of the projected values is the
     average of the middle two (or one) of the projected values of the same two (or one) fixed points (see Figure 3).
    }}
    \label{fig:angular-region-1}
\end{figure}
\noin
\tb{Remarks 2.1}:
The proposition gives a clear foundation for the exact computation of the UF$(\bs{\beta}; F^n_{\mb{Z}})$, or equivalently the PRD($(\bs{\beta}; F^n_{\mb{Z}})=\big(1+\mbox{UF}(\bs{\beta}; F^n_{\mb{Z}})\big)^{-1}$.
\bi
\item[(I)] the  UF$(\bs{\beta}; F^n_{\mb{Z}})$ can be  computed exactly via the optimization over closed sets $\mc{S}_i$. There are unified formulas over $\mc{S}_i$ for distinct cases of permutations. The two types of optimization problems that exist in the proposition are
\bi
\item[(i)]\tb{Type I}: $\min$ (or $\max$) of $\mb{c}'\mb{v}$ for $\mb{v} $ over a closed subset set of $\mbs^{p-1}$ and $\mb{c}\in\R^p$.
\item[(ii)] \tb{Type II}: $\min$ (or $\max$) of $\frac{\mb{b}'\mb{v}}{\mb{v}'\mb{A}\mb{v}}$ for $\mb{v}$ over a closed subset set of $\mbs^{p-1}$ and $\mb{b}\in\R^p$, $\mb{A}\in \R^{p\times p}$ ($A$ could be treated as symmetric and positive-definite over the set).
\ei
\item[(II)]
 $\mb{b}$, $\mb{c}$, and $\mb{A}$ above are determined by $\{\mb{t}_i\}$ and depend on $\mb{v}$ only through the permutation $i_1,\cdots, i_n$ which is induced by the projection of  $\{\mb{t}_i\}$ onto $\mb{v}$. That is, for a given sample and $\bs{\beta}\in \R^p$, and a $\mb{v}\in \mbs^{p-1}$ [or more generally a fixed permutation $i_1,\cdots, i_n$ (of $\{1, 2,\cdots, n\}$) over a set of $\mb{v}$]; $\mb{b}$, $\mb{c}$, and $\mb{A}$ are constant vectors and a matrix.\vs

 Hence, with the constraints discussed in the sequel,  \tb{Type I} optimization can be solved by linear programming and \tb{Type II} optimization can be solved by gradient-type, Newton-type, or interior-point methods (see, e.g. Numerical Recipes (2007) Chapter 10, Freund (2004), and Boyd and Vandenberghe (2004)), among others.
 \hfill \pend
\ei
\noin
\vs
To facilitate the explanation of the basic idea to achieve the exact computation via Proposition 2.1, we first invoke the concept of a ``circular sequence" (see, e.g. Edelsbrunner (1987)).
\vs

Given $n$ general points, $\mb{t}_1,\mb{t}_2,\cdots, \mb{t}_n$ (obtained from ${\mb{Z}}^{(n)}$ and a $\bs{\beta}$) in $\R^p$, and any unit vector $\mb{v}$, assume that $u^{\mb{v}}_{i_1} \leq u^{\mb{v}}_{i_2}\leq \cdots \leq u^{\mb{v}}_{i_n}$ (recall $u^{\mb{v}}_j=\mb{t'}_j\mb{v}$). Then  $\{i_1, i_2, \cdots, i_n\}$ forms a permutation of $\{1,2,\cdots, n\}$ (e.g., see (b) of Fig. 3, where ``4321" represents a permutation from the projection of $4$ points (labeled as $1,\cdots, 4$) to the direction labeled as ``34"). 

\vs
If one rotates $\mb{v}$ counter-clockwise (in $\R^2$), then one will get a sequence of permutations. This periodic sequence of permutations is called a {\it circular sequence} (see the permutations in Fig. 3). In $\R^p$ ($p> 2$), when the unit vector $\mb{v}$ rotates on the unit sphere, we again get a sequence of permutations from the subscripts of ordered projected values, or a circular/spherical sequence.
\vs
\noin
\tb{Some observations on circular/spherical sequences}
\begin{enumerate}
\item[] \tb{O1} The permutation obtained from the projection of the $n$ points
on $\mb{v}$ is exactly the reverse of the permutation obtained from the projection of them
on $-\mb{v}$. 
\item[] \tb{O2} Two successive permutations of a circular/spherical sequence
differ only by switching $p$ integers in the sequence (see (a) of Fig. 2).
\item[] \tb{O3} The permutation changes only whenever the rotation of $\mb{v}$ passes through
a direction perpendicular to a $(p-1)$-dimensional subspace formed by $p$ data points in a given data set that is in general position (defined later)  (see Fig. 3 and (a) of Fig. 2).
\end{enumerate}

\vs
\noin
\tb{Proposition 2.2}: Assume \tb{(A1)} and \tb{(A2)} hold. Let $V\subset \mbs^{p-1}$ be a piece of a unit circle/sphere such that $\forall~ \mb{v}\in V$, $u^{\mb{v}}_{j_1}\leq u^{\mb{v}}_{j_2}\leq \cdots \leq u^{\mb{v}}_{j_n}$. That is, over $V$, $j_1,j_2,\cdots, j_n$ is a fixed permutation of $\{1,2,\cdots, n\}$. Then (i) $N^{-}_{\mb{v}}$ is a constant over $V$; (ii)
there are no $\mb{v}_i\in V ~~(i=1,2)$ such that $\mb{v}_1\neq \mb{v}_2$ and $\mb{v}_i \in S_i$.
\vs
\noindent
\tb{Proof:}\vs
(i) 
When $\mb{v}$ moves over $V$, in order for $N^{-}_{\mb{v}}$  to change its value, it is obvious that
at least one  $k^{\mb{v}}_i$ changes from less than zero to greater than or equal to zero. That is, $\mb{v}$ must cross a $\mb{v_0}$ such that $k^{\mb{v_0}}_i=0$. The latter happens with probability zero under \tb{(A2)}.
\vs

(ii) Assume that there is a $\mb{v}\in S_2\cap V$, then $N^{-}_{\mb{v}}$ is either $0$ or $n$. By (i) there exists no $\mb{v_1}\in V$ such that $\mb{v_1}\in S_1$, since the latter means $0<N^{-}_{\mb{v_1}}<n$, a contradiction. That is, $V\subset S_2$. Similarly, if there is a $\mb{v}\in S_1\cap V$, one can conclude that
$V\subset S_1$.   
\hfill \pend
\vs
To get the exact value of the UF$(\bs{\beta}; F^n_{\mb{Z}})$ utilizing Proposition 2.1, 
 it seems that one has to know the set $\mc{S}_i$ first, $i=1,2$ (or more accurately their boundaries).   $\mc{S}_2$ can be empty. In fact, when the convex hull formed by all $\mb{t}_i$s contains the origin, then $\mc{S}_1=\mbs^{p-1}$. Fortunately, we do not have to identify $\mc{S}_i$, $i=1,2$.  \vs

Since there is no unique formula over $\mc{S}_i$ in the Proposition,  the exact computation task requires us to further partition $\mc{S}_i$ into disjoint pieces. For example, we could partition $\mc{S}_1$ into five pieces and $\mc{S}_2$ into two pieces, according to the cases listed in Proposition 2.1.  The latter task is not as easy as identifying $\mc{S}_i$. 
    For example, identifying all $\mb{v} \in\mc{S}_1$ such that $k^{\mb{v}}_{(n1)}>0$ for an even $n$ case is not straightforward at all. We seek other approaches below.\vs

For a given sample $\mb{Z}^{(n)}$ and a $\bs{\beta} \in \R^p$ and a $\mb{v} \in \mbs^{p-1}$, there is a unique permutation $i_1,\cdots, i_n$ of $\{1,2,\cdots, n\}$ induced by $k^{\mb{v}}_i=1/ \mb{t}'_i \mb{v}$. The  $k^{\mb{v}}_{i_j}$ ($j=1,\cdots, n$) is all we need for the calculation in (\ref{UF-3.eqn}) or Proposition 2.1.  However, a permutation $i_1,\cdots, i_n$ corresponds to a set of $\mb{v}\in\mbs^{p-1}$, with each member of the set capable of  producing the same permutation via $k^{\mb{v}}_i$.
\vs
That is, a fixed permutation corresponds to a unique piece of $\mbs^{p-1}$ (or of the surface of the unit sphere). There are in total
at most  $n!$ possible permutations hence $n!$ disjoint pieces that partition the $\mbs^{p-1}$ (or the surface of the unit sphere).
By Proposition 2.2, each piece  belongs to either $\mc{S}_1$ or $\mc{S}_2$.
Selecting one $\mb{v}$ from each piece is sufficient for the exact computation of the UF$(\bs{\beta}; F^n_Z)$ via Proposition 2.1.
The cost is approximately of order $O(n^{n+1/2})$ without counting optimization cost, which is  computationally unaffordable. We seek to merge some pieces. 
\vs

 In light of Observation 3 (O3) on $\mb{v}$ induced permutations (in regards to the circular or spherical sequence), when $\mb{v}$ moves on the surface of the unit sphere, its induced permutation changes only when it crosses a hyperplane ($H_0$) that goes through the origin and is perpendicular to another hyperplane ($H_1$) that is formed by 
 sample points from $\{\mb{t}_i\}$.
 \vs
  The former hyperplanes ($H_0$'s) (each containing the origin) cut  the $\mbs^{p-1}$ into disjoint $N(n,p)$ pieces $P_k$  ($k=1,\cdots, N(n, p)$), where $N(n, p):= 2\sum_{i=0}^{p-1}{q-1\choose i}$ (see Winder(1966)) and $q:=N_n^p(\{\mb{t}_i\})$ is the total number of distinct $(p-1)$-dimensional hyperplanes formed by 
  points from $\{\mb{t}_i\}$. $q \leq {n \choose p}$. Assume $q> 1$. When $\{\mb{t}_i\}$ are in a general position (IGP) (see Z19a for definition), $q={n \choose p}$. In the latter case, $N(n, p)=O(n^{p(p-1)})$, lower than the cost $O(n^{n+1/2})$ above if $n\geq p (p-1)$.
  \vs

  Each $P_k$ ($k=1,\cdots, N(n,p)$) corresponds to a unique permutation $\{i_1,\cdots, i_n\}$, that is, $1/\mb{t}'_{i_1}{\mb{v}_0}\leq 1/\mb{t}'_{i_2}{\mb{v}_0}\leq \cdots \leq 1/\mb{t}'_{i_n}{\mb{v}_0}$, $\forall~ \mb{v}_0\in P_k$. The latter in turn corresponds to  a polyhedral cone which is determined by
 \be
   \mb{B}'\mb{v}\leq \mb{0}_{(n-1)\times 1}, \label{B.eqn}
 \ee
 where $\mb{v}\in \mbs^{p-1}$ and $B=(B_1,\cdots, B_{n-1})_{p \times (n-1)}$, $B_j:= \mb{t}_{i_j}-\mb{t}_{i_{j+1}}$, $j=1,\cdots, N^{-}_{\mb{v}_0}$; $B_j:= -(\mb{t}_{i_j}-\mb{t}_{i_{j+1}})$, $j=N^{-}_{\mb{v}_0}+1,\cdots, (n-1)$, with the vector inequality in the coordinate-wise sense.
\vs
By Proposition 2.2, the entire $P_k$ belongs to only one  $\mc{S}_i$. So as long as we have one $\mb{v}_0$ from each $P_k$, we can easily produce the permutation associated with the $P_k$ and the induced $k^{\mb{v_0}}_i$ and determine which $\mc{S}_i$ and formulae should be used in Proposition 2.1. Coupled with the constraints $B'\mb{v}\leq \mb{0}_{(n-1)\times 1}$ above, both Type I and Type II optimization problems in the Proposition could be solved in linear time (note that $\mb{b},\mb{c}$ and $\mb{A}$ are constants over the entire piece of $P_k$). The exact computation of the UF$(\bs{\beta}; F^{n}_{\mb{Z}})$ can be achieved with the worst-case time complexity of order $TC(n,p,N_{iter}):=O( N(n,p)(p^{3}+n\log n+np^{1.5}+ npN_{iter})$, where $N_{iter}$ is the number of iterations needed when solving the type II optimization problem.
\vs
\noin

\vs
\begin{figure}[t]
\centering
    \begin{subfigure}[t]{0.47\textwidth}
        \includegraphics[width=\textwidth]{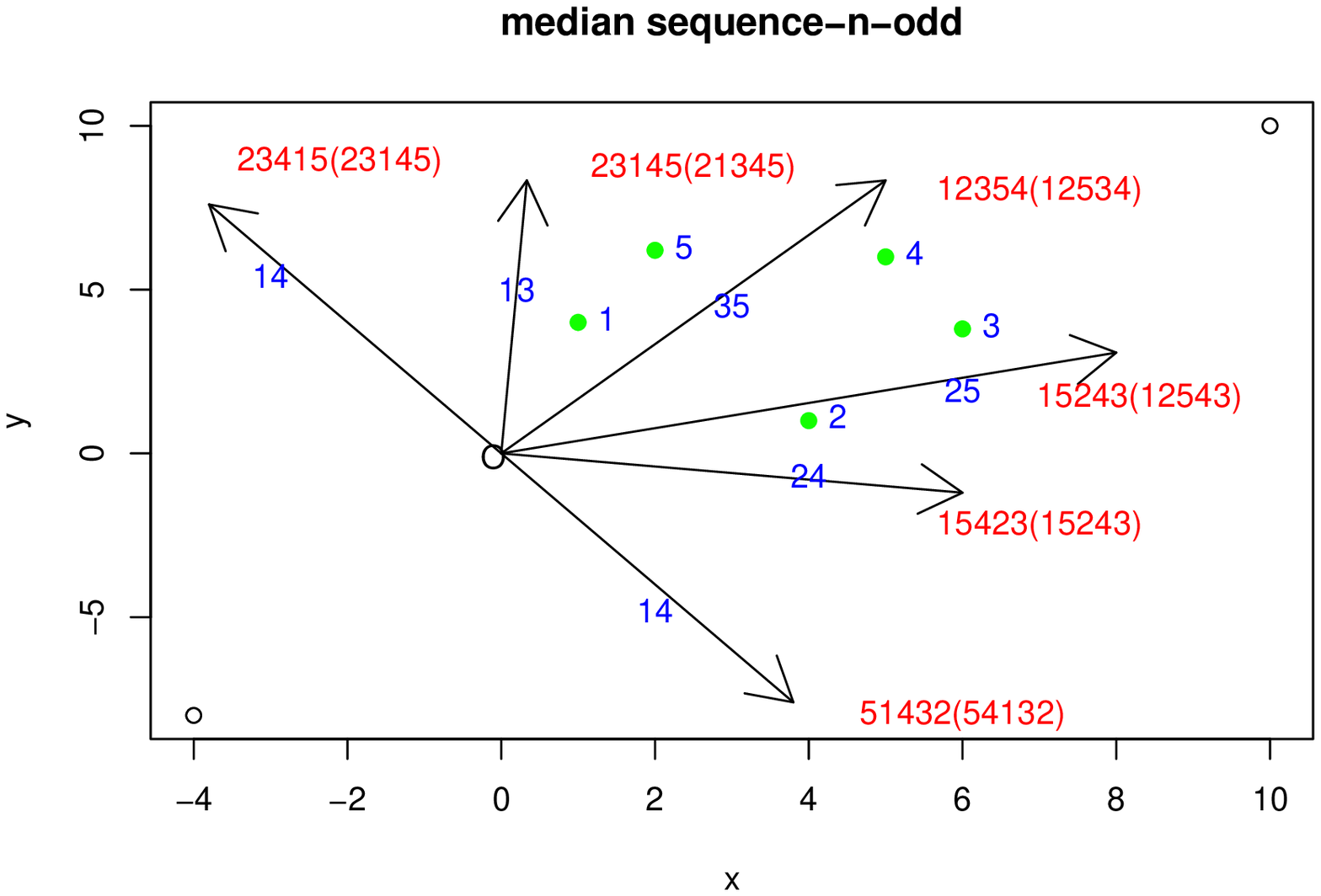}
        \caption{odd-n median sequence demonstration}
        \label{fig:odd-1}
    \end{subfigure}
    \begin{subfigure}[t]{0.47\textwidth}
        \includegraphics[width=\textwidth]{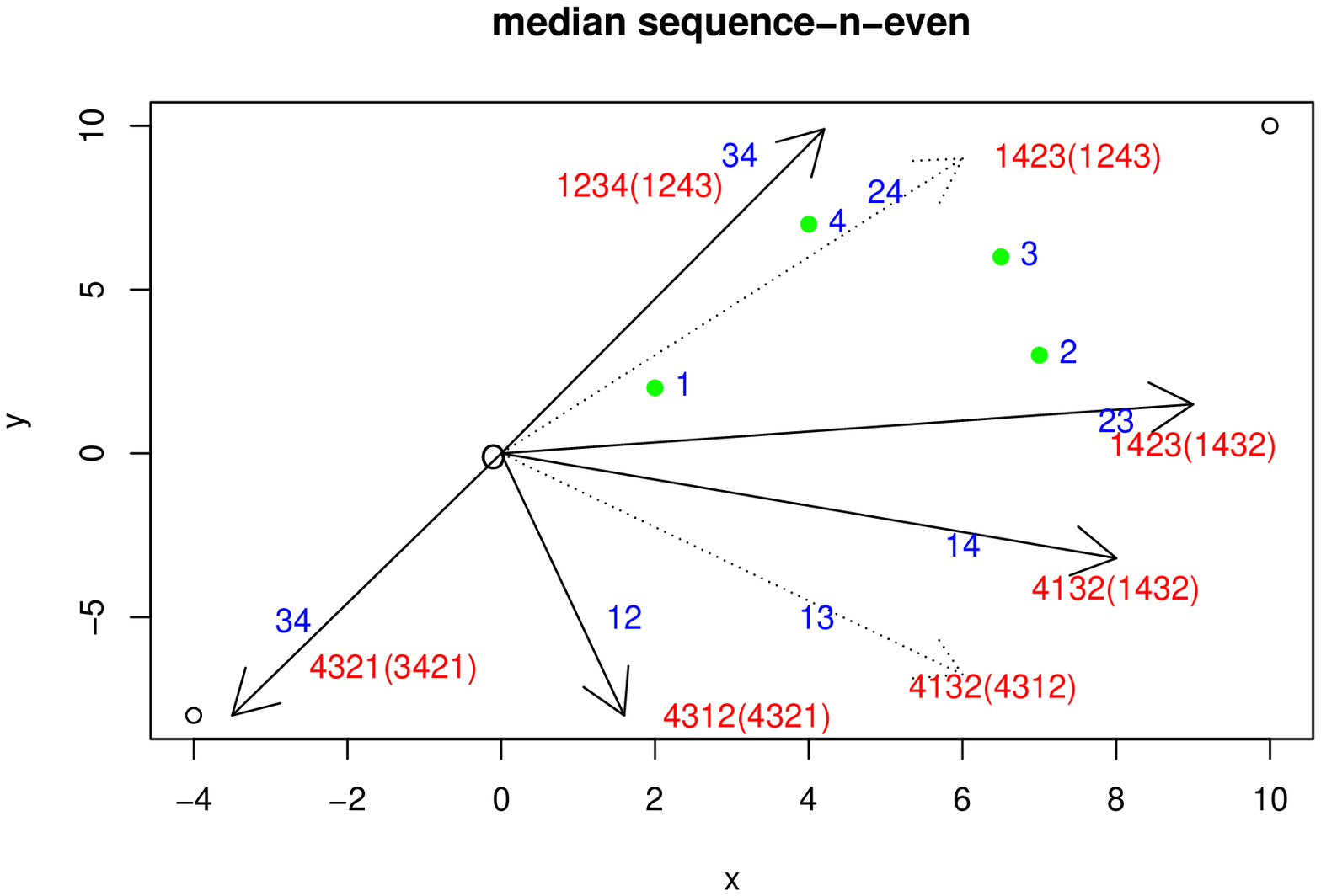}
        \caption{even-n median sequence demonstration}
        \label{fig:even-1}
    \end{subfigure}
        \caption{{
        Median-sequence demonstration. (a)  Five sample points are labeled as ``1",...,``5". Line ``14" cuts the space into two halfspaces. Focusing on the upper right one suffices.
         Label ``ij" means that the labeled ray is perpendicular to the line segment connecting $i$ and $j$.
          When $\mb{v}$ rotates within the angular region formed by ``ij" and ``ik" (or ``kj", or ``jk"),
        the median of the projected values
        is the projected value of the repeated label (point) $i$ (or $j$, or $k$). The median sequence is ``14", ``13", ``35", ``25", ``24" (and ``14"). (b) Four sample points are labeled as ``1",..., ``4". Line ``34" cuts the space into two halfspaces, focusing on the lower right one suffices.
         Along each ray, there are two permutations listed (as in (a)), due to the overlaps of the projected values of some of the two points. The labels of the common middle two points in the permutations help to identify the median sequence  ``34", ``23", ``14", ``12" (and ``34")  which form 4 regions corresponding to two middle point pairs ``4-2" (formed by ``34" (upward), ``23" and O), ``4-3", ``1-3", and ``3-2".}}
        \label{fig:median-sequence-1}
\end{figure}

\noindent
\tb{Theorem 2.1}  Under \tb{(A1)-(A2)}, for a given sample $\mb{Z}^{(n)}$ and a $\bs{\beta}$ in $\R^p$, the UF$(\bs{\beta};F^n_{\mb{Z}})$ (or the PRD$(\bs{\beta}; F^n_{\mb{Z}})$)
can be computed exactly with the worst-case complexity of  (i) $O(n^2\log(n))$  for $p= 2$; 
(ii) $TC(n,p,N_{iter})$ for $p>2$.
\vs

\noindent
\tb{Proof:} For simplicity of description, we assume that  $N_n^p(\{\mb{t}_i\})={n\choose p}$. The general case (with a smaller $N_n^p(\{\mb{t}_i\})$)  could be treated similarly.
\vs
\noin
(i) Consider \tb{the case  $p=2$}. That is, the $\mb{t}_i$ are bivariate points. We show that we can divide the entire
circle $\|\mb{v}\|=1$ into $O(n)$ pieces (arcs) using the so-called median sequence (Zuo and Lai (2011)). These $O(n)$ pieces of arcs further help to divide the entire unit disk into $O(n)$ pieces (each formed by the origin, two radii and a piece of arc) (see (b) of Fig. \ref{fig:angular-region-1}). Over each piece, the middle two numbers (see (b) of Fig. \ref{fig:median-sequence-1}) (or one in the odd n case, see (a) of Fig. \ref{fig:median-sequence-1}) of the projected values $\mb{t'}_i\mb{v}$ are the projected values of some two (or one) fixed points (or point) from $\{\mb{t}_i\}$. 
\vs
In (a) of Fig. \ref{fig:median-sequence-1}, when $\mb{v}$ rotates over the angular region formed by O, if we consider rays labeled as ``ij" and ``ik" (or ``kj", or ``jk")  then the point labeled as ``i" (or ``j") (i.e. the common label) is the single point whose projected value will always be the median of the projected values. The median sequence is the rays ``14" (up), ``13", ``35", ``25", ``24", ``14" (down) which form 5 angular regions corresponding to point ``1" (formed by ``14" (up), ``13",and O) ``3", ``5",``2", and ``4"; whereas in (b) of Fig. \ref{fig:median-sequence-1}, along various rays labeled as ``ij", there are
 permutations listed (also in (a)). Each ray corresponds to two equivalent permutations, because along each direction (ray), the projection of some two points overlap. These permutations help to identify the middle two points and the median sequence. The median sequence is the rays ``34" (up), ``23", ``14", ``12", ``34" (down) which form 4 regions corresponding to two-point pairs ``4-2" (formed by ``34" (upward) ``23" and O), ``4-3", ``1-3",``3-2".
 When $\mb{v}$ rotates over the angular region formed by O, 23, and 34 (up); the points  ``4" and ``2" are the two points whose projected values are the middle two of all projected values (they appear in the middle of the permutations along the rays ``34" (up),``24" ``23").
\vs
\noindent
Figure \ref{fig:median-sequence-1} just illustrates a general  phenomenon with concrete examples. We have generally
\vs
\noin
\tb{Lemma 2.1:} (i) For $p=2$, there are $O(n)$ rays that divide the unit disk into $O(n)$ pieces (cones, or angular regions) $A_j$, each with the origin as its vertex.
Over $A_j$, the median of the projected values $\{\mb{t'}_i\mb{v}\}$ is the projected values of some two (or one in the odd n case) fixed points $\mb{t}_{j_1}$ and $\mb{t}_{j_2}$. (ii) the
UF$(\bs{\beta};F^n_{\mb{Z}})$ and the PRD$(\bs{\beta};F^n_{\mb{Z}})$ can be computed exactly in $O(n^2\log n)$.
\vs
\noindent
\tb{Proof:}~ \vs
 For simplicity, label sample points $\{\mb{t}_i\}$ as $1, 2,\cdots, n$. For each $i$ there are $j_{1}, \cdots, j_{i_k}$ labels (or points), such that the line  connecting $i$ to $j_{m}$ ($1\leq m\leq i_k$), labeled as $``ij_m"$, cuts the plane into two closed halfplanes so that each contains no less than $\lfloor(n+1)/2\rfloor$ points. In  Fig. \ref{fig:median-sequence-1},  $i_k$  is 2 (for odd $n$) and 3 (for even $n$). But they could be larger in other cases.

\vs
Identify the unit vector over the unit circle that is perpendicular to the line $ij_m$
 by its polar coordinate angle $\theta_{ij_m}$ ($0\leq \theta_{ij_m} \leq \pi$) (only halfplane suffices).
 For each $i$, keep the two unit vectors that have the minimum and maximum polar angles, respectively.  In total, there are $O(n)$ such unit vectors. 
 These $O(n)$  vectors cut the unit disk into $O(n)$ angular regions each  formed by the origin and two  unit vectors.  By the construction (also see Fig. \ref{fig:median-sequence-1}), it is readily seen that over each angular region $A_j$,
  the middle two (or one in odd $n$ case, skip mentioning this case hereafter) integers of the permutations are the same.
  When $\mb{v}$ rotates over each region $A_j$ 
  the middle two of the projected values $u^{\mb{v}}_i$ are the projected values of some two fixed points (say, $\mb{t}_{j_1}$, $\mb{t}_{j_2}$).
 This completes the proof of the first part of the Lemma.
 \vs

 Over each piece $A_j$ (in total $O(n)$ pieces), invoking  
 Proposition 2.1 and optimization programming (considering the boundary directions suffices),
 the job can be done with the cost of $O(n^2)$. However, to find out the boundaries of the $O(n)$  pieces, it costs
 $O(n^2\log n)$.  
Thus,  we have the second part of the Lemma.
\hfill \pend
\vs
\noin
(ii) Consider \tb{the cases $p>2$}.
\vs
 Before proving this case, we introduce some basic concepts about a convex body. For more details, refer to Fukuda (2004).
 A hyperplane $H$ of $\R^p$ is \emph{supporting} $P$ (a  p-polyhedron or p-polytope) if one of the two closed halfspaces of H
contains $P$. A subset $F$ of $P$ is called a \emph{face} of $P$ if it is either $\varnothing$, $P$ itself, or the intersection
of $P$ with a supporting hyperplane. The faces of dimension $0, 1, \mbox{dim}(P)-2~
\mbox{and}~ \mbox{dim}(P) -1$ are called the \emph{vertices}, \emph{edges},
\emph{ridges} and \emph{facets}, respectively.
\bee
\mbox{A hyperplane}~ H &=& \{\mb{x} \in \R^p ~| \mb{a}'\mb{ x} =c \}, \mb{a} \in \R^p \setminus \{\mb{0}\}, c \in \R^1,\\
\mbox{A closed halfspace}~ H &=& \{\mb{x} \in \R^p ~|\mb{a}'\mb{x} \leq c\},\\
\mbox{A polyhedron}~ P& =& \{\mb{x} \in \R^p ~| A\mb{x} \leq \mb{b}\}, A \in \R^{m\times p},\mb{b} \in \R^m,\\
\mbox{A Polytope}~ P &=& \{\mb{x} \in \R^p ~| A\mb{x}\leq \mb{b}, \mb{I} \leq \mb{x} \leq \mb{u} \}, \mb{I},\mb{u} \in \R^p,\\
\mbox{A polyhedral cone} ~P &=& \{\mb{x} \in \R^p ~| A\mb{x} \leq \mb{0}\},
\ene
\vs
Obviously, exact computation is achieved if we can obtain the RHS of display (\ref{UF-4.eqn}). For the latter, we appeal to Proposition 2.1.
To implement the proposition, 
we need to solve the two types of optimization problem (see Section 2.4 for implementation).\vs
By the discussion immediately before the theorem, we know the key for the optimization problems is to identify all pieces  $P_k$ ($k=1,\cdots, N(n,p)$) of $\mbs^{p-1}$. Equivalently, we need to identify all $N(n,p)$ distinct permutations of $\{1,2,\cdots, n\}$. The latter is equivalent to finding a unit vector $\mb{u}\in P_k$ for each $P_k$ which can produce the unique fixed permutation over $P_k$.
\vs
Each $P_k$ is the intersection of $\mbs^{p-1}$ and the
polyhedron cone formed by the constraint $B'\mb{v}\leq \mb{0}_{(n-1)\times 1}$ in (\ref{B.eqn}). The edge (or ridge) of the cone can be used to find the $\mb{u}$ above, which is shared by another adjacent cone. In other words, the edge (or ridge) is the intersection of (at least) two hyperplanes $H_0$s which go through the origin and are perpendicular to two hyperplanes $H_1$s, each of which is formed by  points from $\{\mb{t}_i\}$, respectively.
\vs
The direction from the origin to any other point on the intersection hyperline of two hyperplanes $H_0$s is a solution of the vector sought above.
We denote the direction by $\mb{u}$ ($\mb{u}$ could also be obtained more costly via the origin and any vertex of a polytope through vertex enumeration,
  see Bremner et al., 1998, Paindaveine and \' Siman (2012) and Liu and Zuo (2014)).\vs
Each $\mb{u}$ lies on the boundary of $P_k$. It not only lies in the facet of one cone but also
 lies in that of an adjacent cone which shares the common intersection hyperline (edge or ridge) with the former cone (cf. Fig. 1 of Mosler et al (2009)). A tiny perturbation
 of $\mb{u}$ in opposite directions will cause $\mb{u}$ to enter the interiors of the two adjacent cones. There might be more than two cones that are adjacent. Thus, every $\mb{u}$ might yield two or more new permutations (the scheme in the algorithm yields up to $8 \times(p-2)$ distinct ones, $p> 2$). \vs

 Update the total number $N_{permu}$ of distinct permutations. 
With respect to each distinct permutation, or equivalent over each $P_k$, update $\sup_{v\in\mbs^{p-1}}|g(\mb{v})|$ according to Proposition 2.1 and carry out one of the two types of optimization.\vs

Repeat above steps (find more $\mb{u}s$) until $N_{permu}=N(n,p)$ or the UF can not be improved after trying $\kappa p$ more distinct permutations ($\kappa$  is a positive integer, which could be something like 30).

\vs
For a given data set (or  $\{\mb{t}_i\}$), the total number of $H_1$s is fixed, but for each $H_1$, there are infinitely many $H_0$s ($p>2$) which go through the origin and are perpendicular to $H_1$. So by utilizing different $H_0$s  one can always obtain all distinct permutations in theory.
If one obtains $N(n, p)$ distinct permutations which means that each piece of $P_k$ has been visited (or all relevant directions $\mb{u}$s have been obtained), the resulting UF (or PRD) is exact in theory. In practice, however, not every distinct permutation updates the  UF.
In the latter case, the stopping rule ``until UF can not be improved" becomes handy.
 \vs

The cost of computation of key elements of the descriptions above is as follows.
\bi
\item[(a)] calculating all $\{\mb{t}_i\}$ (assume they are in a general position) and  $N(n,p)$  costs $O(np)$,
\item[(b)] calculating normal vectors $\mb{v}_i$ of $H^i_1$, normal vectors $\mb{u}_i$ of $H^i_0$ ($i=1, 2$), and $\mb{u}$ (which is perpendicular to $\mb{u}_i$) costs $O(p^3)$,
\item[(c)] producing each permutation costs $O(n(p+\log n))$,
\item[(d)] updating the total number of distinct permutations cost $O(nN_{permu})$,
\item[(e)] updating $\sup_{v\in\mbs^{p-1}}|g(\mb{v})|$ according to Proposition 2.1 costs
\bi
\item[(i)] $O(n)$ for obtaining $k^{\mb{v}}_i$ ($i= n1,n2$), $\mb{t}_{i_{n1}}$, and $\mb{t}_{i_{n2}}$,
\item[(ii)] $O(p^{1.5}n+p^{2.5})$ for linear programming (see Yin Tat Lee and Aaron Sidford (2015) which is even further improved by Cohen, Lee, and Song (2019)),
\item[(iii)] $O(npN_{iter})$  for the type II non-convex and nonlinear optimization problem. One can use the conjugate gradient method or the even better primal-dual interior-point method (Wright (1997), Morales, et al (2003)) combined with the sequential quadratic programming (SQP) ( Nocedal and Wright (2006)), e.g. package LOQO (Vanderbei and Shanno (1999) and  Vanderbei (1999)).  Where $N_{iter}$ is the number of iterations needed in LOQO.
\ei
\ei
\vs
Keeping only the dominating terms, we have the overall worst-case time complexity $TC(n,p, N_{iter})=O(N(n,p)(p^3+n\log n+np^{1.5}+ npN_{iter}))$.

 This completes the proof of the theorem.
\hfill \pend
\vs

\noin
\tb{Pseudocode} (Exact computation of the UF$(\bs{\beta}; F^n_{\mb{Z}})$, or equivalently of  the PRD$(\bs{\beta}; F^n_{\mb{Z}})$)
\bi
\item Input: Given a sample $\mb{Z}^{(n)}:=\{(\mb{x}'_i, y_i)', ~ i=1,\cdots, n\}$ 
 and a  $\bs{\beta}$ in $\R^p$,
\item Calculate $\{\mb{t}_i\}$  (assume they are in a general position) and $N(n,p)$;
 set UF$= N_{permu}=0$.
\item While ($N_{permu}<N(n,p)$)
\bi
\item[1] Obtain $\mb{u}$ and its induced permutations, store distinct permutations and update its total number $N_{permu}$.
\item[2] Update UF$=\sup_{\mb{v}\in\mbs^{p-1}}|g(\mb{v})|$ via Proposition 2.1 and carry out the corresponding optimization for each distinct permutation.
\item[3] 
 If UF can not be improved after trying $\kappa p$ more distinct permutations ($\kappa=30$), break the loop.
\ei
\item Output: UF (or $1/(1+\mbox{UF})$) of the $\bs{\beta}$ with respect to $F^n_{\mb{Z}}$. \hfill \pend
\ei

\subsubsection{Exact computation algorithms}
\vs
\noin
\tb{(I) Algorithm for the exact computation of the UF$(\bs{\beta};F^n_{\mb{Z}})$ and the PRD$(\bs{\beta};F^n_{\mb{Z}})$ in $\R^2$}
\vs
\noin
Before listing the key steps of the algorithm, we make some comments.
\vs
(i) Directions that are perpendicular to the line segment connecting $\mb{t}_i$ and $\mb{t}_j$
could be the boundary of angular regions, 
so we will have to include them in our calculation.
\vs

(ii) $\sup_{\mb{v}\in \mc{S}_i}|g(\mb{v})|$ ($i=1,2$)  can be obtained  along the median sequence
and the directions given in (i) above.
\vs
\noin
\tb{Exact Algorithm EA-UF2D}
\vs
\noin
 \tb{Input} a $\bs{\beta}$ and $n$ data points $\mb{Z}^{(n)}=\{( \mb{x}_i, y_i)'\}$
 in $\R^2$; \tb{Output}  UF$(\bs{\beta};F^n_{\mb{Z}})$ and PRD$(\bs{\beta};F^n_{\mb{Z}})$.
 \begin{itemize}
 \item []\tb{Initial Step:} (i) Obtain $N:=N_n^p(\{\mb{t}_m\})$ unit vectors $\mb{u}_k(i,j)$ that are perpendicular to the
all possible $N$ hyperplanes formed by $\mb{t}_i$ and $\mb{t}_j$ from $\{\mb{t}_m\}$, $k=1, \cdots,N$.
(ii) Sort $\mb{u}_k(i,j)$ according their polar angles such that $\alpha_{i_1}\leq \alpha_{i_2}\leq\cdots,
\leq \alpha_{i_N}$. (iii) Record the pair $(i,j)$ associated with $\alpha_k$ as the pair $(i^k,j^k)$.

Set a seven-component initial matrix $I_0$ with initial values corresponding to the seven cases in
Proposition 2.1. Let $k=0,  M_k=I_0$, $\mb{u}_k = (1, 0)'$ and $i_1,\cdots i_n$ be a permutation induced by $\mb{u}_k$.
If $i \neq j$ and $\mb{t}'_i\mb{u}_k = \mb{t}'_j\mb{u}_k = \mb{t}'_{i_{kk}}\mb{u}_k$,
set $m^1_k = i^k=i$ and $m^2_k = j^k=j$, else $m^1_k = m^2_k =i^k=j^k= i_{kk}$, where $kk = \lfloor(n + 1)/2\rfloor$.
\vs
 \noin
\item[] \tb{Loop step:} While ($k<=N+1$) $\{$
Let $\mb{v}=\mb{u}_k$.  Update $M_k$ according to Corollary 2.1.
Let $k=k+1$, $\mb{v}=(cos(\alpha_k), sin(\alpha_k))$. If the set $S^m_{k-1}:=\{m^1_{k-1}, m^2_{k-1}\}$ intersects with the set $S_k:=\{i^k,j^k\}$,
then let $m^1_k = i^k, m^2_k = j^k$; otherwise, let $k=k+1$, $m_k^i=m_{k-1}^i$, $i\in \{1,2\}$.
$\}$

For (i in 1:n) $\{$
get $\mb{v}_i$ that is perpendicular to $\mb{t}_i$,
using $\mb{v}_i$ to update $M_{N+i-1}$ according to Proposition 2.1 and to obtain $M_{N+i}$.$\}$\vs
 \noin
\item[]
\tb{Final step:} Set the maximum no-zero element of $M_{N+n}-I_0$ be the UF$(\bs{\beta},F^n_{\mb{Z}})$.
\end{itemize}

\vs
\noin
\tb{(II) Algorithm for the exact computation of UF$(\bs{\beta};F^n_{\mb{Z}})$ and PRD$(\bs{\beta};F^n_{\mb{Z}})$ in $\R^p$, $p>2$}

\vs
\noindent
\tb{Exact Algorithm (EA-UFHD)}
\vs
\noin
 \tb{Input} a $\bs{\beta}$ and $n$ data points $\mb{Z}^{(n)}=\{( \mb{x}'_i, y_i)'\}$
 in $\R^p$; \tb{Output}  UF$(\bs{\beta};F^n_{\mb{Z}})$ and PRD$(\bs{\beta};F^n_{\mb{Z}})$.
 \bi
\item[(a)] Compute $N(n,p)$ and call it by $N$, let 
$k_{pm}=0$.~~~$(O(p)$, assume that $q={n \choose p}$).

\item[(b)] Construct two non-parallel hyperplanes $H_i$ ($i=1, 2$) (each of which is formed by p points from $\{\mb{t}_i\}$) with normal vectors
$\mb{v}_i$, $i=1,2$, respectively.

    ~~~$(O(pn+p^3))$

   Find two hyperplanes $H^{\perp}_i$ that are through the origin and perpendicular to $H_i$  and with normal vectors $\mb{u}_i$, $i=1,2$, respectively.  ~~~$(O(p^3))$

   Let $\mb{U}$ be the unit vectors matrix each of its $(p-2)$ columns is perpendicular to both of $\mb{u}_1$ and $\mb{u}_2$. ~~~ $O(p)$

\item[(c)] For each column vector of the $\mb{U}$  above, call it  $\mb{u}$, introduce eight vectors $\pm \mb{u}\pm\mb{v}_1\pm\mb{v}_2$.
For each of eight  vectors, obtain its induced permutation, and store it in a matrix if it is a new one. Update the total number of
distinct permutations $k_{pm}$. ~~$O( n(p+\log n+ k_{pm}))$

    For each distinct new permutation and the associated vector $\mb{v}$, carry out the optimization  to update $|g(\mb{v})|$
    via Proposition 2.1. 
   Also use $\mb{v}_1$ and $\mb{v}_2$  to update $|g(\mb{v})|$ via Proposition 2.1.
    ~~~$(O((n+\max\{np^{1.5}+p^{2.5}, npN_{iter}\}))$

\item[(d)] while ($k_{pm}<N$)
    $\{$ do (b), (c) above until
         either $k_{pm}=N$ or UF cannot be improved.$\}$ ~~~$(O(N(p^3+(\max\{p^{2.5}+np^{1.5}, npN_{iter}\})+n(p+\log n)))))$

\item[(e)] 
Output the final UF$(\bs{\beta};F^n_{\mb{Z}})$ via Proposition 2.1 and (\ref{UF-4.eqn}).~~~$(O(1))$

 \ei\vs
Overall cost in the worst case is $O(N(n,p)(p^3+n\log n+np^{1.5}+ npN_{iter}))$.\vs

\vs
\noin
\tb{Remarks 2.2}\vs
\tb{(I)} EA-UFHD exactly follows the idea given in the proof of Theorem 2.1. Since there are infinitely many $H^{\perp}_i$s (that go through the origin) for each $H_i$,
in theory one can get all distinct permutations, equivalently all desirable unit directions $\mb{u}$, and the final UF (or PRD) obtained is in theory exact. \vs
\tb{(II)} In the best scenario, $N(n,p)$ can be replaced by  $O(n^2)$. 
Even in this case, the cost of exact computation in the worst case is $O(n^2(p^3+n\log n+np^{1.5}+ npN_{iter})$, which is still unaffordable for large $n$ and/or $p$. An approximate algorithm, such as \tb{AA-UF-3} (introduced below) with cost of order $O(N(np+p^3))$, where tuning parameter $N$ being the total number of normal directions of the hyperplanes formed by p points from $\{\mb{t}_i\}$, is more feasible in practice. \vs

\tb{(III)} On the other hand, besides theoretical interest itself, without the slow exact algorithm as the benchmark, no one can develop fast practically feasible approximate algorithm with known acceptable accuracy for large $n$ and $p$. The contribution and importance of the exact algorithm can never be over-emphasized.
\hfill  \pend

\vs
However, exact computation of the UF$(\bs{\beta}; F^n_{\mb{Z}})$ or the PRD $(\bs{\beta}; F^n_{\mb{Z}})$ is not our primary goal. Our ultimate goal is to seek
depth induced median or other estimators.  Practically, the latter has to be computed approximately. In the following, we discuss some more practically feasible approximate algorithms.
\vs

\subsection{Approximate computation}

 Approximate computation of statistical depth functions  is common and has been carried out in Rousseeuw and Struyf (1998), Dyckerhoff (2004), Cuesta-Albertos and Nieto-Reyes (2008), Chen et al. (2013), and Zuo (2018) and in the references cited therein.
\vs

Here we present three approximate algorithms.
The first one is a straightforward naive one. It randomly selects a fixed number $N$ directions from a distribution
(e.g. uniform on the hypersphere), and decorrelating the data before calculation the UF$(\bs{\beta}; F^n)$ defined in (\ref{UF-2.eqn}) along those directions.
\vs
\noindent
\tb{Approximate algorithm AA-UF-1}
\vs
\noin
 \tb{Input} a $\bs{\beta}$ and $n$ data points ${\mb{Z}}^{(n)}=\{( \mb{x}'_i, y_i)'\}$
 in $\R^p$; \tb{Output}  UF$(\bs{\beta};F^n_{\mb{Z}})$ and PRD$(\bs{\beta};F^n_{\mb{Z}})$.
\begin{itemize}
\item[(a)] Randomly select $N$ unit directions $\mb{v}\in\mbs^{p-1}$ according to a uniform distribution on the hypersphere,  use the formula given in (\ref{UF-2.eqn}) or (\ref{UF-1.eqn})   to calculate/update $\sup_{\mb{v}\in \mbs^{p-1}}|g(\mb{v})|$.
\\[1ex]
(Overall cost  is $O(npN)$, the cost to find median can be as low as $O(n)$). \hfill \pend
\end{itemize}
\vs
The second approximate algorithm below employees the idea in EA-UFHD. It considers the directions that represent the edges of the convex cones, where the cones stem from the origin and partition the entire sphere $\mbs^{p-1}$ into disjoint (convex) pieces.
\vs
 When $\mb{v}$ moves over each piece, the permutation induced is fixed. By the fundamental theorem of linear programming, the solution of  the maxima or minima of a linear function over a convex polygonal region occurs at the region's corners.
 (Note that we no longer have linear functions in the \tb{Type II} optimization scenario).
\vs
\noindent
\tb{Approximate algorithm AA-UF-2}
\vs
\noin
 \tb{Input} a $\bs{\beta}$ and $n$ data points ${\mb{Z}}^{(n)}=\{( \mb{x}'_i, y_i)'\}$
 in $\R^p$; \tb{Output}  UF$(\bs{\beta};F^n_{\mb{Z}})$ and PRD$(\bs{\beta};F^n_{\mb{Z}})$.
\begin{itemize}
\item[(a)] Compute the $\mb{t}_i$, $i=1,\cdots, n$. (total cost $O(np)$) 
\item[(b)] Sample two sets $P_i$, each with p points, from $\{\mb{t}_i\}$. Construct two hyperplanes $H_i$ with normal vectors $\mb{v}_i$,  uniquely determined by $P_i$, respectively. Try different $P_2$ until  $\mb{v}_2$ is not parallel to $\mb{v}_1$. (total cost $(O(p^3)$)
\item[(c)] Construct two hyperplanes $H^{\perp}_i$ (with normal vectors $\mb{u_i}$) that is through the origin and perpendicular to $H_i$, respectively. (total cost $(O(p^3))$)
\item[(d)] Obtain $\mb{v}=\mb{u}_1\times \mb{u}_2$ and $\mb{v}_0=\mb{v}/\|\mb{v}\|$; use $\mb{v}_0$ and the formula in (\ref{UF-2.eqn})
  to update $\sup_{\mb{v}\in \mbs^{p-1}}|g(\mb{v})|$. (total cost $O(np)$)
\item[(e)]   Repeat (b)-(d) $N$ times. (total cost $O(N(np+2p^3))$).
Overall cost is $O(N(np+2p^3)+np))$.   \hfill \pend
\end{itemize}
The one below uses  $N$  normal vectors of the hyperplanes determined by $p$ points from
$\{\mb{t}_i\}$.
\vs
\noindent
\tb{Approximate algorithm AA-UF-3}
\vs
\noin
 \tb{Input} a $\bs{\beta}$ and $n$ data points ${\mb{Z}}^{(n)}=\{( \mb{x}'_i, y_i)'\}$
 in $\R^p$; \tb{Output}  UF$(\bs{\beta};F^n_{\mb{Z}})$ and PRD$(\bs{\beta};F^n_{\mb{Z}})$.
\begin{itemize}
\item[(a)]  Compute the $\mb{t}_i$, $i=1,\cdots, n$. (total cost $O(np)$)

\item[(b)] Sample p points from $\{\mb{t}_i\}$, find  the normal vector $\mb{v}$ of the hyperplane determined by them. Along $\mb{v}$,
use the formula (\ref{UF-2.eqn}) to  calculate/update $\sup_{\mb{v}\in \mbs^{p-1}}|g(\mb{v})|$. (total cost $O(p^3+np))$

\item[(c)] Repeat (b) $N$ times. (total cost $O(N(p^3+np))$).~ Overall cost $O(N(np+p^3)+np))$.   \hfill \pend
\end{itemize}
\noin
\subsection{Examples}
\vs
To better understand the algorithms in the last two subsections, we present some examples. Algorithms for the deepest regression lines discussed in Section 3 are also employed below. \vs

 For the exact algorithm, we now explain the implementation of the two types of optimization.\vs
Given a direction $\mb{v}\in P_k$, a permutation, say, $i_1,\cdots, i_n$,  is obtained. That is,  for all the values from $\{k^{\mb{v}}_i=1/\mb{t}'_i\mb{v}\}$,
we have $k^{\mb{v}}_{i_1}\leq k^{\mb{v}}_{i_2}\leq \cdots \leq k^{\mb{v}}_{i_n}$, $\forall ~\mb{v}\in P_k$. The \tb{Type I} optimization problem can be described as
\bi
\item[]\tb{minimize}: $ 
\mb{c}'\mb{v}$, 
\item[]\tb{subject to}: (i) $\mb{B}'\mb{v}\leq \mb{0}_{(n-1)\times 1}$; ~~  (ii) $\mb{v}'\mb{v}=1$,
\ei
where $\mb{c}$ is a constant vector,  $\mb{B}$ is a constant matrix, and minimization could also be exchanged for maximization. That is, we have a linear objective function with a linear inequality constraint and a quadratic equality constraint.\vs
When $p=2$, each $P_k$ becomes a piece of an arc of the unit circle and the cones formed by the linear constraints are the angular regions with two radii as their boundaries. The optimization problem becomes
linear programming over the piece of arc. By the fundamental theory of linear programming, the minimization
or maximization occurs only at the boundary. Therefore, only evaluation of $\mb{c}'\mb{v}$ is needed for $\mb{v}$ at the two boundary directions. There are at most $O(n^2)$ pieces of $P_k$'s. \vs Generally,
the \tb{Type I} optimization problem can be solved by an augmented
Lagrangian minimization using the R package `alabama’, or by sequential quadratic programming using the
R solver `slsqp’. Alternatively, it can be 
transformed into semidefinite programming
problems and solved using the R solver `csdp’. Also the R packages `optisolve' and `nlopt' are applicable. \vs

Now we turn to the \tb{Type II} optimization problem. It can be described as
\bi
\item[]\tb{minimize}: $ 
\frac{\mb{b}'\mb{v}}{\mb{v}'\mb{A}\mb{v}}$, 
\item[]\tb{subject to}: (i) $\mb{B}'\mb{v}\leq \mb{0}_{(n-1)\times 1}$; ~~  (ii) $\mb{v}'\mb{v}=1$,
\vs
where $\mb{b}$ is a  constant vector,  $\mb{A}$ 
and $\mb{B}$ are constant  matrices, $\mb{A}$ can be treated as a symmetric and positive definite one, and minimization can also be exchanged for maximization.
\ei
\vs That is, we have a non-linear, non-convex, but differentiable objective function or a rational objective function with a linear inequality constraint and a quadratic equality constraint.
The problem again can be solved by using the R packages `alabama’,  `optisolve',  and `nlopt'.

\vs
\noin
\tb{Example 2.4.1~~Performance of exact and approximate algorithm}.~~ Here we examine the performance of the exact versus the approximate algorithm (EA-UF2D v.s. AA-UF-1) for computing the UF, w.r.t. their accuracy,  speed, and estimated mean squared errors.
\vspace*{-3mm}
\bec
\begin{figure}[h]
    \centering
    \begin{subfigure}[t]{0.47\textwidth}
        \includegraphics[width=\textwidth]{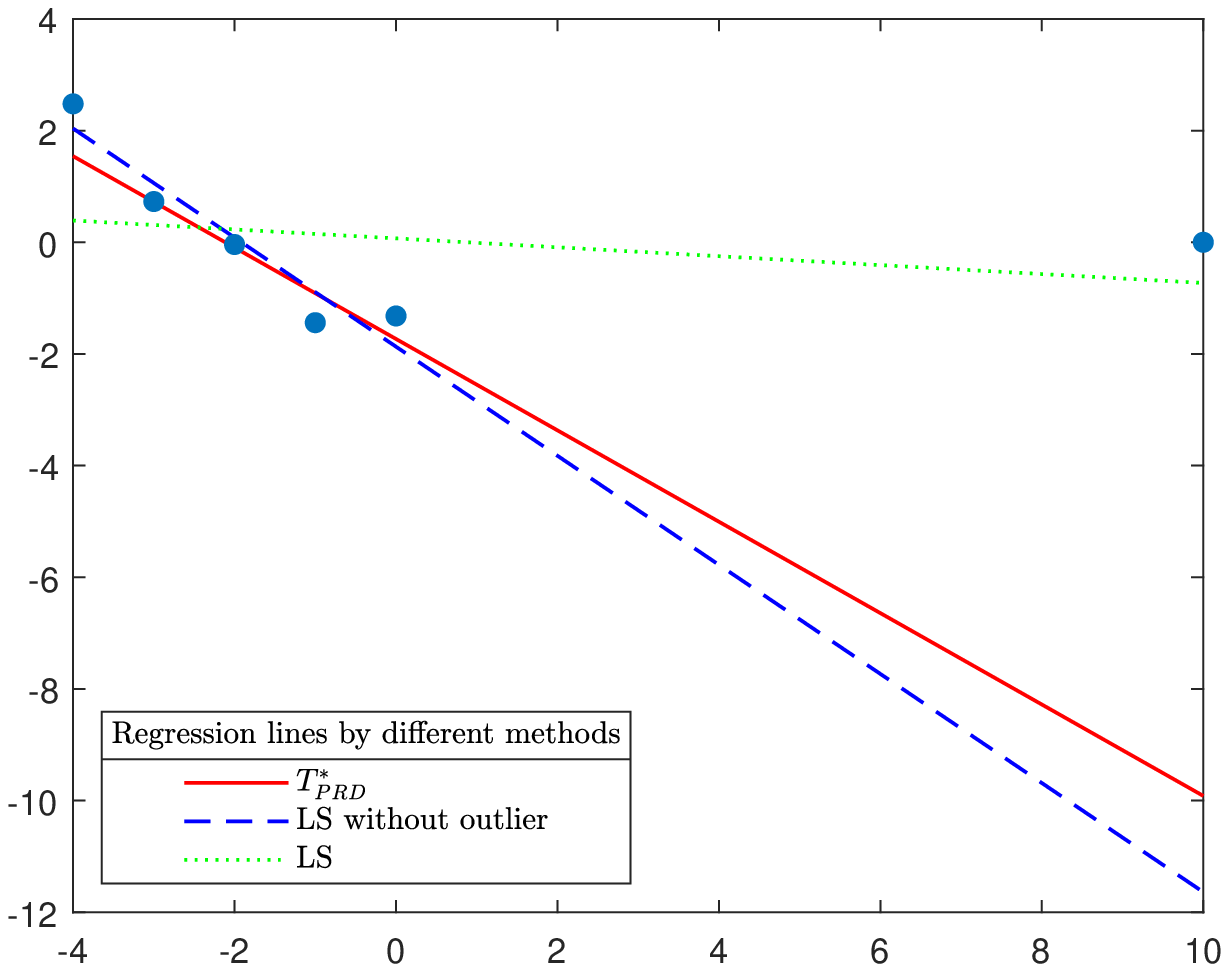}
        \caption{three lines and six points}
        \label{fig:six-points}
    \end{subfigure}
    \begin{subfigure}[t]{0.47\textwidth}
        \includegraphics[width=\textwidth]{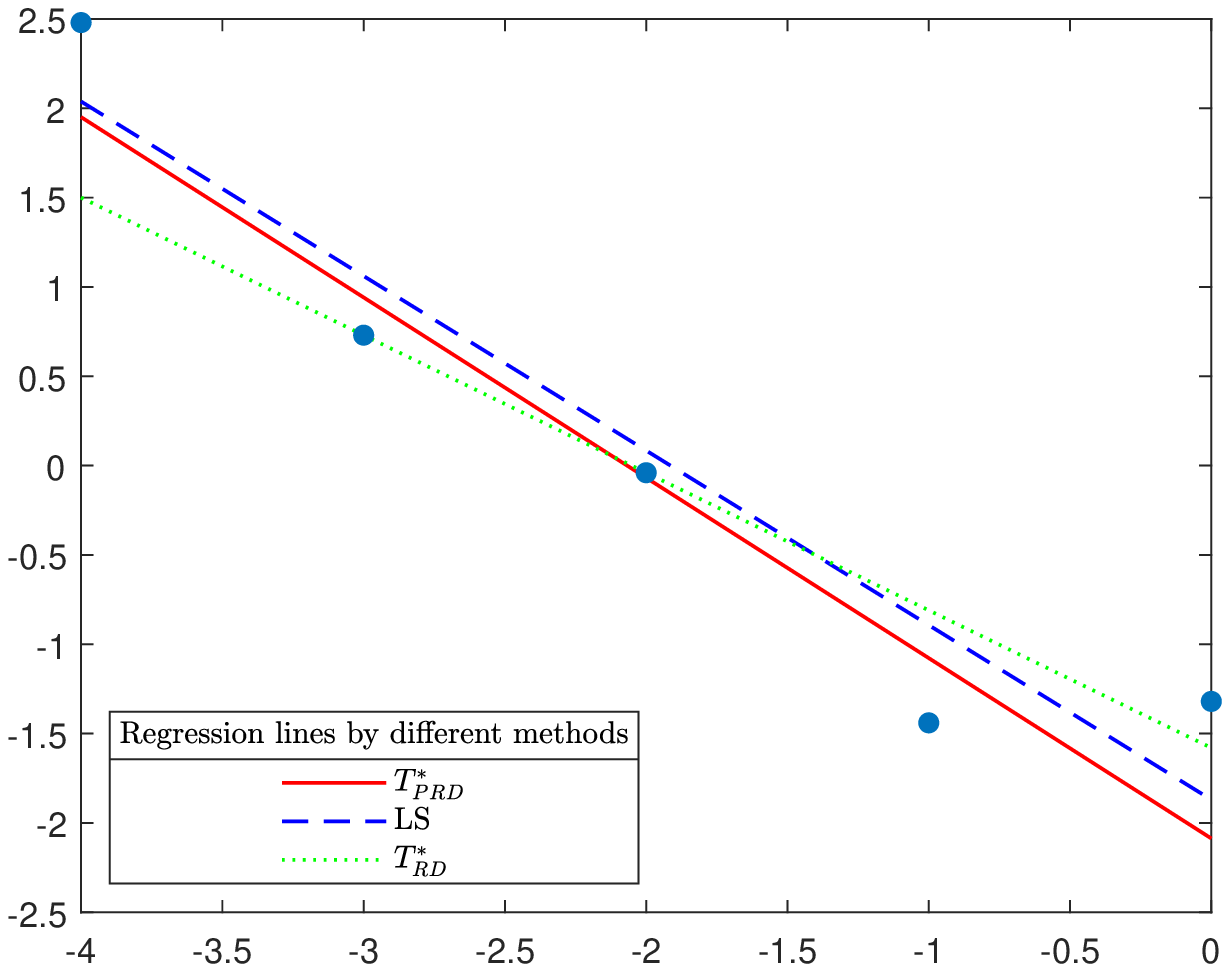}
        \caption{three lines and five points}
        \label{fig:five-points}
    \end{subfigure}
    \caption{ {
     (a) Solid red: the deepest regression line induced from the PRD. Dashed blue: the least squares
     line based on five points without the horizontal outlier. Dotted green: the least squares line based on
     all six points.
     (b) Solid red: the deepest regression line induced from the PRD w.r.t five points. Dashed blue: the
     least squares line based on five points without the horizontal outlier. Dotted green: the deepest regression line induced from the RD of RH99 w.r.t. five points.
    }}
   \label{fig:three-lines}
   \vspace*{-5mm}
\end{figure}
\enc
 For illustration purposes, we utilize the data set given in Huber and Ronchetti (2009) Exhibit 7.1. Three regression lines are obtained w.r.t. the data. The first line  $T^*_{PRD}$ is the maximum  PRD line ($\bs{\beta_1}=(-1.7317456,-0.8184845)'$ in (intercept, slope)' format); 
 the second one (LS without outlier) is the least squares line \emph{without} the horizontal outlier ($\bs{\beta_2}=(-1.87, -0.977)'$); the third one (LS) is the least squares line ($\bs{\beta_3}=(0.07, -0.08)'$) w.r.t. all points. See (a) of Figure \ref{fig:three-lines}.
 \vs
 Next, we calculate the unfitness of  the three lines ($\bs{\beta}'s$).
 First, the unfitness, reported in Table 1, is calculated without the horizontal outlier  for the fairness of the comparison between exact and approximate algorithms. That is, they are calculated w.r.t. just five points. 
 \vs

 Second, using all six points, the results are very similar to those in Table 1 and details are omitted. As an example, the  UF from the EA are $0.8340, 1.2113,$ and $2.3367$ and the UF from the AA (mean of 1000 replications) are
 $0.5246$, $0.6855$ and $2.1846$, respectively.
 \vs
Consistent with expectations, $\bs{\beta}_1$ ($T^*_{PRD}$) has the lowest UF,  $\bs{\beta}_2$ has the second lowest UF, and the $\bs{\beta}_3$ has the highest UF. That is, \textit{in terms of the UF  ordering}, $T^*_{PRD}$ is the best choice among the three while $\bs{\beta}_3$ is the worst, fitting with the intuitive comprehension of (a) of Figure \ref{fig:three-lines}.
\vs
 At the same time, it is not difficult to determine the regression depths (RD) of RH99 of the three lines, they are $2/6$, $1/6$, $1/6$, respectively. (For simple methods of calculation of the RD, see RH99 or RS98).
 That is, \emph{in terms of the $RD$ ordering}, the least square line $\bs{\beta}_3$ is as deep (or good) as the line $\bs{\beta}_2$,
 while both are less deep than the PRD induced line $T^*_{PRD}$, which is somewhat inconsistent with the intuitive comprehension of (a) in  Figure \ref{fig:three-lines}.
 Of course, the comparison here is  not very fair since the different methods (PRD vs RD) are based on different objective criteria and $\bs{\beta}_2$  and $\bs{\beta}_3$  use a different number of total points. 
\vs
In (b) of Figure \ref{fig:three-lines}, all three lines are calculated w.r.t. just five points without the outlier. The solid red line is the deepest  line induced from the PRD with $\bs{\beta_1}=(-2.083114,-1.009444)'$,
the dashed blue line is the same as in (a), the dotted green line is the deepest  line  from the RD of RH99
with $\bs{\beta_3}=(-1.58, -0.77)'$. The RDs of the three lines are 2/5, 1/5, 3/5, respectively.
This time, as expected from $T^*_{RD}$, the line induced from the RD becomes the deepest one.
\vs
\vspace*{-5mm}
\bec
\begin{table}[h!]
\centering
Table entries (a,b,c,d) are a:= mean of UF,  b:=standard deviation (sd) of UF, c:=time consumed (in seconds), d:=number of unit vectors used.\\[2ex]
\begin{tabular}{c c c c} 
\!\!\!&\!\!\! $\bs{\beta}_1$ (line $L_1$) &\!\!\!\!$\bs{\beta}_2$ (line $L_2$) &\!\!\! \!\!$\bs{\beta}_3$ (line $L_3$)\\[1ex]
\hline
\!\!\!EA& \!\!\!(0.59, 0, 1.11e-3, 13) &\!\!\!\!(0.87, 0, 1.28e-3, 15) &\!\!\!\!\! (2.88, 0, 1.30e-3, 15) \\[1ex]
\!\!\!AA & \!\!\!(0.58,  4.1e-3, 1.657e-2, e+3) &\!\!\!\!(0.86,  4.8e-3, 1.662e-2, e+3) &\!\!\!\!\! (2.85,  3.4e-2, 1.664e-2, e+3) \\[1ex]
\hline
\end{tabular}
\caption{
~Performance of exact and approximate algorithms w.r.t. different $\bs{\beta}'s$ (lines).}
\label{ex-aa.tab}
\end{table}
\enc
\vspace*{-5mm}
In Table 1, the calculation of the approximate algorithm (AA) is repeated  1000 times to mitigate the randomness; the mean and the standard deviation of 1000 UFs are calculated. 1000 unit vectors are used in the calculation per replication, and the time consumed per replication is reported in the table.
\vs
 The table reveals that the exact algorithm (EA) is  much 
  faster than the AA. The former used  no more than 15 unit vectors whereas the latter, using 1000 vectors, still returned a smaller (under-approximated) UF than the exact one.  This is the beauty of the EA. If the number of unit vectors used is
increased to $10^4$, then the UF from the AA is still smaller than the one from the EA which just employed 13 unit vectors, in the $\bs{\beta_1}$ case. Since there is no fluctuation in the EA, all the sd's are zero. The time reported is the average of per replication from $1000$ replications.  Observations above hold only in this special ($n=5$ and $p=2$) example. Note that when $n$ and/or $p$ increase, the EA is no longer feasible in practice.\hfill \pend
\vs
\noin
Results 
above and below in this section are all based on Matlab {code}
 {running} on a desktop: Intel(R)Core(TM) i7-2600 CPU @ 3.40GHz. All Matlab as well as R and C++ code in this article are downloadable via
 https://github.com/zuo-github/comp-prd-medians.
 \vs

\vs
\noin
\tb{Example 2.4.2~~Performance of exact algorithm versus approximate algorithms}\vs  In the last example, the dimension $p=2$ and $n=5$ are too limited in practice. In this example, we consider standard normal random points with $p=3$ and $n=10$,  For this fixed small data set, we employ the EA and the AAs
to compare their performance. Here we seek the best output from them, that is the largest UF (note that  the UF is defined based on the
supremum of univariate unfitness; so generally speaking, the larger the better).\vs
For the EA, the stopping rule of the algorithm is
the total number of distinct permutations ($N_{perm}$) used by it whereas for {the AAs}, the stopping rule is the total number of
unit directions ($N_{\mb{v}}$) employed by the AAs. The outputs from different algorithms are listed in {the} table below. Column one is the number of distinct permutations
used by the EA-UFHD, the second column is the number of  while loops that need to be executed to get the desired total distinct permutations. The fourth column is the total number of directions used by the AAs.
\vs
\bec
\begin{table}[h!]
\centering
\begin{tabular} {c c c c c c c}
& \underline{EA-UFHD } &  & & &\underline{AAs }  &\\[1.5ex]
$N_{permu}$ & $N_{while-loop}$ &UF & $N_{\mb{v}}$ used & AA-UF-1 & AA-UF-2 &AA-UF-3\\[1ex]
used&executed&&& \underline{UF} & \underline{UF} & \underline{UF} \\[1ex]
\hline\\[.5ex]
120 &18 &4.3505&120&4.0414&4.3862&5.4190\\[1ex]
200 &35 &5.1988&1,000&5.8134&5.2446&5.4190\\[1ex]
300 &61& 6.3708&10,000&6.6369&6.4582&5.4190\\[1ex]
400 & 92& 8.2669&100,000& 6.7699&6.6985&5.4190\\[1ex]
500 & 136& 8.2669&1,000,000&6.7924&6.8182&5.4190\\[1ex]
600  & 187   &8.2669 &10,000,000&6.8930&6.8675 &5.4190\\[1ex]
\hline\\[.5ex]
\end{tabular}
 \caption{
 ~Performance comparison of exact versus approximate algorithms on a fixed standard normal data set with the size $3$ by $10$.} \label{table:2}
\end{table}
\enc
\vspace*{-8mm}

Inspecting the table   immediately reveals that (i) in order to use $120$ distinct permutations, one needs to run the while loop $18$ times in the EA and can get $4.3505$ for the UF. Note that  each single while loop yields a $p$ by $(p-2)$ unit vector matrix $\mb{u}$ (in this case a single column vector) and each
column of $\mb{u}$ will induce eight unit vectors (see (c) of the EA-UFHD), these eight vectors  will lead to at most eight distinct permutations in each while loop. This explains why only 18 while loops {are} needed.
The number $120$ {(equal to ${10 \choose 3}$)} in this case is exactly the total possible planes  formed by any three sample points from a data set (IGP) of size $10$ in a three dimensional space.   (ii) if one uses $400$ distinct permutations, then $92$ while loops {are} needed  and one can get the $8.2669$ for the UF, which is the final answer for {the} exact UF.  (iii) for the AAs, the AA-UF-3 yields the same UF for all {number} of directions considered, as long as the latter is no less than $120$. The meaning of the latter number is explained above.
 This phenomenon  is {not} surprising and  it is due to
the design of the algorithm which only utilizes the normal vectors of all possible planes formed by three points, the latter  is a fixed number $120$ in this case. (iv) the AA-UF-1 yields larger UF than that of AA-UF-2 in most cases. (v) most important, the AAs even after exhausting $10$ million directions still can not get a comparable size of UF that the exact algorithm produces {using just} $400$ distinct permutations.
\vs
Notice that in this example, the number of total possible distinct permutations is $N(n,p)=2*({119 \choose 0}+{119 \choose 1}+{119 \choose 2})= 14282$.
However,  $400$ distinct permutations are enough for the exact UF. The superiority of {the} EA over {the} AAs is clearly demonstrated.
In summary, the AAs can run much faster than the EA, but it is very difficult for them to get the results the EA provides (in fact, AA-UF-1 yields the UF $6.8930$ even after depleting $10$ million directions).\vs On the other hand,
the EA is not feasible in practice for larger $n$ and $p$, I would not recommend it for $p>3$. Recall, the possible distinct permutations in  $p=3$ and $n=10$ case above is $14282$. In table \ref{table:2}, we used $400$ permutations to get the exact result of the UF leading immediately to an issue.
That is, in practice what should be the cut off number $N$ for the distinct permutations{?} One suggestion is $N=\min\{{n\choose p}+500, 1000\}$
(obviously, there is an alternative automatic stopping rule based on the UF) and the while loop should be executed at most $ 1000$ times. \hfill \pend
\vs
However, there are at least two drawbacks in the comparison above, (i) the data set is still small, (ii) all results are  random (since all use random sampling) (which means perhaps $200$ permutations are enough for the exact UF if the EA runs many times. Similarly, the AA using $1000$ directions might get much better results if it runs many times). Each result in table \ref{table:2} comes from a single run. Therefore the results and conclusions above have their limitations.\vs
\vs
\noindent
\tb{Example 2.4.3 Performance of exact algorithm versus approximate algorithms}\vs In this example,
we will consider the cases $p=3, n=50$ and $p=4, n=20$. Furthermore, we will run each of  the algorithms $100$ times (ideally it should be $1000$ or even $10,000$ times, but due to the time consumed by the EA-UFHD, they are not affordable options). The $q$ for the two cases are $19600$ and $4845$ and $N(n, p)$ are $384140402 $ and $37887089270$, respectively.
\bec
\begin{table}[t!]
\begin{tabular}{ c c c c c c}
\hline\\[.5ex]
dimension& $N_{\mb{v}}$ used &methods  &UF(mean, min, max)& $D_{max}$ & $N_{nega}$\\[1ex]
$p=3$ &   400&            EA&  (2.0797    1.8820    2.2315) &        0  &       0\\[1ex]
$n=50$ &      &      AA1    & (2.0242    1.8475    2.1118) &  -0.1197  & 74.0000\\[1ex]
           &       &      AA2     & (1.9094    1.7287    2.0822) &  -0.1493  & 95.0000\\[1ex]
           &&             AA3      & (2.0800    1.9811    2.1335) &  -0.0981  & 50.0000 \\[2ex]
   & 800 & EA&     (2.2673    2.0406    2.4942)&         0  &       0\\[1ex]
   &     & AA1&     (2.2217    2.0031    2.3756)&   -0.1186&   58.0000\\[1ex]
   &     & AA2&     (2.1308    1.8936    2.3436)&   -0.1506 &  82.0000\\[1ex]
   &     & AA3&    (2.3037    2.1754    2.3762)&   -0.1180 &  36.0000 \\[3ex]
$p=4$ & 360 &  EA&   (3.3443    2.7657    4.6322) &        0 &        0\\[1ex]
 $n=20$&  &   AA1&  (2.8832    2.4389    3.4670)&   -1.1652&   91.0000\\[1ex]
 &&            AA2&  (2.5424    2.1256    3.1373) &  -1.4949&   98.0000\\[1ex]
 &&            AA3&  (3.4406    3.1707    3.5357) &  -1.0964&   36.0000\\[2ex]
 & 540& EA&     (3.8939    2.8989    6.0121)&         0 &        0\\[1ex]
  && AA1&  (2.9247    2.3428    4.4621)&   -1.5500&   93.0000\\[1ex]
  && AA2&  (2.1350    1.4178    3.2886)&   -2.7235&  100.0000\\[1ex]
  && AA3&  (4.6775    4.1511    4.7682)&   -1.2439&   11.0000\\[1ex]
 \hline\\[.5ex]
\end{tabular}
 \caption{
 ~Performance of the EA and the AAs with respect to different data sets and directions used.}
 \label{table-example-2.4.3}
\end{table}
\enc
\vspace*{-10mm}
\vs
Furthermore, we try to set up a fair comparison base for the two types of methods. We will use the number of directions employed by them as the criterion for {a} performance assessment. Note that for the EA, we count the number of directions it utilized as follows. In each while loop, there are 8*(p-2) directions induced by the $\mb{u}$ matrix; furthermore, there are $\mb{v}_1$ and $\mb{v}_2$ vectors, so {in total} $2+8*(p-2)$ unit vectors are employed by the EA.
\vs
We (i) compute the mean, minimum (min) and maximum (max) of 100 UF{s} for each of the algorithms, and (ii) compute $D_{max}:=$ the difference in max UF (the max of UF of any algorithm {subtracted} by that of {the} EA) and (iii) count the times that the UF of any algorithm is less than that of the EA in the $100$ trials (denote by $N_{nega}:=$ the count of the negative $D_{uf}$s, $D_{uf}$  is defined as the UF of any algorithm subtracted by that of the EA for each of  $100$ trials).
\vs
 Results are listed {in} table \ref{table-example-2.4.3}.  Note that in the $p=3$ case, $400$ and $800$ directions used by the {AA} amounts to $40$ and $80$ while loops executed in the EA-UFHD (we convert them to the number of while loops in EA to terminate the algorithm). In the $p=4$ case, $360$ and $540$ directions for the {AA} amounts to $20$ and $30$ while loops executed in the EA. \vs

Examining the table reveals that (i) in all cases considered, the EA produces the largest UF among the $100$ trials, this is demonstrated by the
max or $D_{max}$ column in the table; (ii) the AA3 is the second best method among all four in the sense that (a) it yields the largest UF among the three AA methods in mean, minimum, and maximum of  $100$ UFs in all cases, and (b) its mean and minimum of $100$ UFs are even larger than those of the EA in all cases considered
(except the most important maximum of  $100$ UFs), and (iii) the last columns in table \ref{table-example-2.4.3} {shows} the percentage that the EA  yields a larger UF than that of any other AAs. For example,
in the $p=4$, $n=20$, and $N_{\mb{v}}=540$ case, the UF of the EA is larger than that of AA2 in every trial. \hfill \pend

\vs
\vspace*{-1mm}
\bec
\begin{table}[h!]
\centering
Performance of approximate algorithms w.r.t.  efficiency and accuracy\\[1ex]
\begin{tabular}{c c c c c c} 
p&methods & mean UF &standard deviation  &time consumed& number of $\mb{v}$ used\\[.5ex]
\hline\\[.0ex]
2&AA-UF-1&0.3035  &0.1053  &26.4750 & $10^3$ \\[.5ex]
&AA-UF-2 & 0.3041 &0.1054 &116.0503 & $10^3$\\[.5ex]
&AA-UF-3 &0.3042  &0.1057 & 34.7244 & $10^3$\\[1ex]
5&AA-UF-1&0.4815 &0.0994  &31.1470& $10^3$\\[.5ex]
&AA-UF-2 &0.4447  &0.0995 &227.3421 & $10^3$\\[.5ex]
&AA-UF-3 &0.5472  &0.1095 &80.3334 & $10^3$\\[1ex]
10&AA-UF-1&0.5198 & 0.0928 &33.4609& $10^3$\\[.5ex]
&AA-UF-2 &0.4467 &0.1049 &278.6988 & $10^3$\\[.5ex]
&AA-UF-3 &0.7385  &0.1156 &100.4025  & $10^3$\\[1ex]
20&AA-UF-1&0.5253  &0.0877  &40.5291 & $10^3$\\[.5ex]
&AA-UF-2 &0.4152  &0.1054  &570.2156 & $10^3$\\[.5ex]
&AA-UF-3 & 1.1656  &0.1626&236.5488  & $10^3$\\[1ex]
\hline
\end{tabular}
\caption{
~Performance comparison of the three approximate algorithms.}
\label{aa-aa.tab}
\end{table}
\vspace*{-10mm}
\enc
{The examples above} clearly demonstrate the advantage of the EA over the AAs. But as said before, the EA is not very feasible in practice due to its limitation on $p$ and $n$. For example, it is not capable of handling the case $p=20$ and $n=100$ and running $1000$ times with
{the} while loop executed $20$ times in each trial. But the AAs are more feasible and can easily handle the situation.  Furthermore, the depth induced median has to be computed eventually by approximate methods. The following example is devoted to
the comparison of the three AAs.\vs
\vs
\noin
{\bf Example 2.4.4 Performance comparison between three approximate algorithms} 
\vs
 Here we generate $m=1,000$ samples from the  model: $y_i=\beta_0+\beta_1 x_{i_1}+\cdots +\beta_{p-1}x_{i_{p-1}}+e_i, i=1,2,\cdots, n,$
with sample sizes $n=100$, where $e_i \sim N(0,1)$. In light of the regression equivariance of {the} deepest projection depth estimator and the invariance of {the} PRD, we can assume without loss of generality (w.l.o.g.) that  $\bs{\beta}_0=(\beta_0, \beta_1,\cdots, \beta_{p-1})'=(0,0,\cdots, 0)'$. We generate $\mb{Z}_i=(x_{i_1},\cdots, x_{i_{p-1}}, y_i)$ from {a} $p$-dimensional standard normal distribution. 

\vs
{The three} AAs compute the unfitness of $\bs{\beta}_0=(0,0,\cdots, 0)'$ with results (means and deviations of 1000 UFs, \tb{total time} consumed (in seconds) for $1000$ samples, and unit directions used) listed in Table \ref{aa-aa.tab} which features the attributes of the three AAs.
\vs

The table reveals that (i) the AA-UF-1 is fastest and the AA-UF-2 is slowest in all cases, confirming the theoretical time complexity results;
(ii)the AA-UF-3 is the most accurate in all cases (with the largest mean UF), and the AA-UF-2 is superior over the AA-UF-1 \emph{only for the case $p=2$} in terms of accuracy (mean is slightly larger); (iii) the AA-UF-1 is most efficient (smallest s.d. and fastest), and the AA-UF-3 has the largest s.d., but this  could be reversed by increasing the number of directions $\mb{v}$ used  to $n \choose p$; (vi) p has the least (no greater than linear) effect on the time consumed by AA-UF-1 and it reduces the s.d. of AA-UF-1
when it increases; (v) overall, the AA-UF-3 (or the AA-UF-1) {is} recommended. \hfill \pend

\section{Computation of the PRD induced median}

\subsection{Algorithms}
Our fundamental goal for introducing the notion of depth in regression is not the depth itself but its induced median (and other estimators).
Computation of the latter is our eventual objective. It seems that there is no analytical way to identify the depth induced median.
Consequently the computation of the median in high dimensions has to be carried out approximately.\vs

Before addressing the approximate computation of the maximum projection depth estimator (or median), we first show that it indeed deserves to be called a median since it recovers the univariate sample median when $p=1$.
Recall that (assume (w.l.o.g.) again $S_y=1$)
\be
\bs{\beta}^*_{PRD}=\arg\min_{\bs{\beta}\in \R^p}\sup_{\mb{v}\in\mbs^{p-1}}\Big|\mbox{Med}_i\{\frac{y_i-\mb{w}'_i\bs{\beta}}{\mb{w}'_i\mb{v}}\} \Big|.
\label{beta*.eqn}
\ee
When $p=1$, it reduces to the following
\be
{\beta}^*_{PRD}=\arg\min_{{\beta}\in \R}\sup_{{v}=\pm 1}\Big|\mbox{Med}_i\{\frac{y_i-{\beta}}{{v}}\} \Big|. \label{beta-cover-median.eqn}
\ee
 We have\vs
\noin
\tb{Proposition 3.1} 
When $p=1$, 
the ${\beta}^*_{PRD}$ recovers  the regular sample median of $\{y_i\}$.\vs
\noin
\tb{Proof}:
~
Let $y_{(1)}\leq y_{(2)}\leq \cdots \leq y_{(n)}$ be the ordered values of $\{y_i\}$ and $\mu=(y_{(n1)}+y_{(n2)})/2$, where $n1$ and $n2$ are defined in Proposition 3.1, i.e., $\mu$ is the regular sample median of $\{y_i\}$. We show that $\beta^*_{PRD}=\mu$. 
It is readily seen that 
\be
{\beta}^*_{PRD}=\arg\min_{{\beta}\in \R}\big|\mbox{Med}_i\{{y_i-{\beta}}\}\big|=
\arg\min_{{\beta}\in \R}\big|\mbox{Med}_i\{y_i\}-{\beta}\big|= \arg\min_{{\beta}\in \R}\big|\mu-{\beta}\big|,
 \label{proof-cover-medain.eqn}
\ee
where the first equality follows from (\ref{beta-cover-median.eqn}) and the oddness of the median operator; the second one follows from the translation equivalence (see page 249 of RL87 for definition) of the median as a location estimator;
the third one follows from the definition of $\mu$.\vs
The RHS of (\ref{proof-cover-medain.eqn}) above indicates that $\mu$ is the only solution for ${\beta}^*_{PRD}$.
\hfill \pend
\vs
\noin
\tb{Remark 3.1}\vs
 The proposition holds true for the univariate population median. That is, ${\beta}^*_{PRD}$ also recovers the univariate median in the population case.
\hfill \pend
\vs

Now we turn to the approximate computation of $\bs{\beta}^*_{PRD}$ in (\ref{beta*.eqn}). First, we notice that the $\bs{\beta}^*_{PRD}$ must be bounded, or equivalently, the search for the optimal $\bs{\beta}$ in the RHS of (\ref{beta*.eqn}) can be limited within a bounded set (hypersphere). \vs
\noindent
\tb{Proposition 3.2}. For a given $\mb{Z}^{(n)}=\{(\mb{x}'_i, y_i), i=1,\cdots, n\}$ and a $\bs{\beta}\in \R^p$, under \tb{(A1)} there is a constant $c^*$ such that
$$
\bs{\beta}^*_{PRD}=\arg\min_{\|\bs{\beta}\|\leq c^*}\sup_{\mb{v}\in\mbs^{p-1}}\Big|\mbox{Med}_i\Big\{\frac{y_i-\mb{w}'_i\bs{\beta}}{\mb{w}'_i\mb{v}}\Big\}\Big|.$$

\noin
\tb{Proof}:
To see this, notice that for a given $\bs{\beta}\neq 0$, let $\mb{v}_0=\bs{\beta}/\|\bs{\beta}\|$, then
\bee
\mbox{UF}(\bs{\beta}; F^n_Z)&=& \sup_{\mb{v} \in\mbs^{p-1}}\Big| \mbox{Med}_i\{(y_i-\mb{w}_i'\bs{\beta})/\mb{w}'_i\mb{v}\}\Big| 
\geq \Big| \mbox{Med}_i\{(y_i-\mb{w}_i'\bs{\beta})/\mb{w}'_i\mb{v}_0\}\Big|\\[1ex]
&=& \Big| \mbox{Med}_i\{y_i/\mb{w}'_i\mb{v}_0\}- \|\bs{\beta}\|\Big| 
\longrightarrow \infty,~a.s., \mbox{~~~~as $\|\bs{\beta}\|\to \infty$,}
\ene
where the last step follows from the fact that $\mbox{Med}_i\{y_i/\mb{w}'_i\mb{v}_0\}$ is bounded a.s.. 
 Let $\delta=\mbox{UF}(\mb{0},F^n_Z)$ and $c^*=\sup\{\|\bs{\beta}\|, \mbox{UF}(\bs{\beta}, F^n_Z)\leq \delta\}$. This completes the proof. \hfill \pend
\vs
\noin
The rough idea of the approximate algorithm is as follows.
Randomly select $N_{\bs{\beta}}$ of  $\bs{\beta}$ over a very \emph{wide} range in the parameter space $\R^p$, calculate all UF$(\bs{\beta},F^n_{\mb{Z}})$ w.r.t. the sample distribution $F^n_{\mb{Z}}$ of $F_{\mb{Z}}$. Sort the latter and select $p+1$ $\bs{\beta}$s with the smallest unfitness. Over the simplex formed by these $p+1$ $\bs{\beta}$ points (in the parameter space), search the point ($\bs{\beta}$) with the smallest unfitness (equivalent to the deepest regression line or hyperplane). Denote the latter by $\mb{T}^*_n$, the sample version of the $\mb{T}^*_{PRD}$.
\vs
\noin
In the above process, we have implicitly taken  advantage of the property of the PRD$(\bs{\beta};F_{\mb{Z}})$ or the UF$(\bs{\beta};F_{\mb{Z}})$.
That is, the PRD$(\bs{\beta};F_{\mb{Z}})$ satisfies the property (P3) of Z18a (monotonicity relative to the deepest point). Therefore, the depth region
of $\bs{\beta}$ (the set of all $\bs{\beta}$s with  depth no less than a fixed value)  is convex and nested. Hence, the deepest point(s) must lie over the closed convex simplex formed by the $p+1$ $\bs{\beta}$ points. The deepest PRD point is unique (see Zuo (2019b)).
\vs
The following is an {\textbf{approximate algorithm}} for the computation of $\mb{T}^*_n$.
\begin{itemize}
\item[] (A) Randomly select a set of points $\bs{\beta_j}\in\R^p$, $j=1,\cdots, N_{\bs{\beta}}$, where $N_{\bs{\beta}}$ is a tuning parameter for the  total number of  random points $\bs{\beta}$ tried.
\item[] (B) For each $\bs{\beta_j}$,  compute,  over a set of randomly selected unit directions $\mb{v_k}\in\mbs^{p-1}, k=1,\cdots, N_{\mb{v}}$, an approximate unfitness of $\bs{\beta_j}$ w.r.t.  $\{ Z^j_{ik} =(y_i-\mb{w_i'}\bs{\beta_j})\big/ (\mb{w_i'}\mb{v_k})\}$,  $i=1,\cdots, n$, $k=1,\cdots, N_{\mb{v}}$,  where $N_{\mb{v}}$ is another tuning parameter.

\item[] (C) 
Select the deepest $p+1$ $\bs{\beta}_j$s (points with the smallest unfitness).
Search over the closed convex hull formed by these $p+1$ points via a common nonlinear optimization algorithm (e.g. the downhill simplex method (Nelder-Mead), or the MCMC technique) to get the final deepest $\bs{\beta}$ or our approximate $\mb{T}^*_n$.
\item[] (D) To mitigate the effect of randomness, repeat the steps above (many times) so that the one  $\mb{T}^*_n$ with the maximum updated regression depth is adopted.
\end{itemize}
\vs
\tb{Remarks 3.2}:
 \begin{itemize}
 \item[(I)]  The candidate  (random point) $\bs{\beta}$ can be produced by randomly selecting $p$ points from $\mb{Z}^{(n)}=\{ (\mb{x_i}, y_i)', i=1,\cdots, n\}$ which determine a ($\bs{\beta}$ or)  hyperplane $y=\mb{w'}\bs{\beta}$ containing all $p$ points. Let $S_{\bs{\beta}}:=\{\bs{\beta}_1,\cdots, \bs{\beta}_{N_{\bs{\beta}}}\}$ be all $\bs{\beta}$s.
 \vs

\item[(II)] 
The random directions could be selected among those which are normal vectors of the hyperplanes  formed by $p$ points  from $\mb{Z}^{(n)}$ above. Furthermore, for each $\bs{\beta}_j \in S_{\bs{\beta}}$, one can consider all $\bs{v}^j_i=(\bs{\beta_i}-\bs{\beta_j})/{\|\bs{\beta_i}-\bs{\beta_j}\|},~\forall~ \bs{\beta}_i\neq \bs{\beta}_j$.
    Let $S_{\bs{v}}:=\{\bs{v}_1,\cdots, {\bs{v}}_{N_{\bs{v}}}\}$ be all $\bs{v}$s.
    \vs

\item[(III)] For a better approximation of depth (unfitness) of $\bs{\beta}_j$, tune (increase) $N_{\mb{v}}$. For a better approximation of $\mb{T}^*_n$, tune
$N_{\bs{\beta}}$.
Continue iterating until it satisfies a stopping rule (e.g. the difference between consecutive depths is less than a cutoff value).\vs

\item[(IV)] The overall worst case time complexity of the algorithm is: step (A)+(B): $O(pN_{\bs{\beta}}(p^2+nN_{\mb{v}}))$,
 step (C):  $O(pN_{\bs{\beta}}+N_{\mb{v}}N_{Iter}pn)$, where 
  $N_{Iter}$ is the total number of iterations in the optimization algorithm, step(D) $O(RpN_{\bs{\beta}}(p^2+nN_{\mb{v}}N_{Iter}))$, where $R$ is the number of replications. The overall cost of the
    algorithm is $O(Rp N_{\bs{\beta}}(p^2+nN_{\mb{v}}N_{Iter}))$, which could be reduced to $O(pN_{\bs{\beta}}(p^2+nN_{\mb{v}}))$
  by skipping steps (C) and (D).
\vs
\item[(V)] After obtaining the approximate UF of the first $(p+1)$ $\bs{\beta}_j$s, record $\mbox{UF}_{min}$, the minimum of all $(p+1)$ UFs. For the calculation of the UF for any future $\bs{\beta}_k$, if along any direction $\mb{v}$, the directional $\mbox{UF}_{\mb{v}}(\bs{\beta}_k, F^n_{\mb{Z}})$ \mbox{(see (13) of Z18a)}$\geq \mbox{UF}_{min}$, then stop the computation for $\bs{\beta}_k$ and move to $\bs{\beta}_{k+1}$.  Update $\mbox{UF}_{min}$ if a new UF is obtained. In this way, the overall cost of the algorithm will be drastically reduced.
\vs
\item[(VI)] \tb{Alternative algorithms}.\vs (i) After (A), compute the coordinate-wise median of the $\bs{\beta}$s and use it as an initial point for a nonlinear optimization algorithm (e.g. optimx or DEoptim in R) along with other arguments (e,g. a function compute-UF)  
 to find the $\mb{T}^*_n$.
  \vs(ii) Increase $N_{\mb{v}}$ and $N_{\bs{\beta}}$ and skip steps (C) and (D). Namely, just employ steps (A)+(B). \vs
 (iii) Seek an algorithm similar to MEDSWEEP in Van Aelst et al (2002) for the $\mb{T}^*_{RD}$ since when $p=1$ and regression through the origin, the $\mb{T}^*_{PRD}$ recovers the median of $\{y_i/x_i\}$ (see  Section 3.3 of RH99 for a related discussion).
 \hfill \pend
\end{itemize}

\subsection{Computation times}
\noin
As requested, we now turn to the comparison of the computation times for the $\mb{T}^*_{PRD}$, the $\mb{T}^*_{RD}$ and the least trimmed squares (Rousseeuw (1984)) regression (ltsReg) estimator. \vs
As a median in regression,  $\mb{T}^*_{{RD}}$ is a promising robust alternative to the classic least squares (LS) regression estimator. In fact,
in terms of asymptotic breakdown point (ABP) robustness, the former possesses a $33\%$ ABP (Van Aelst and Rousseeuw (2000) (VAR00)), in contrast to the $0\%$ of the latter. \vs
Zuo (2019b) (Z19b) has investigated the ABP of $\mb{T}^*_{PRD}$, it turns out that it possesses the highest possible ABP, $50\%$. For this advantage over
the $\bs{\beta}^*_{{RD}}$ (see illustration examples in Z19b), it has to pay a price in the computation. The cost of the computation of  the $\bs{\beta}^*_{{RD}}$ is generally lower than that of the $\bs{\beta}^*_{PRD}$.
 \vs
 To see the difference in the computation behavior, we list in table \ref{table-comp-time} the computation time consumed by both medians for different sample size $n$s and dimension $p$s. For the benchmark and comparison purposes, we also list the times consumed by the famous least trimmed squares
(Rousseeuw (1984)) regression (ltsReg) estimator. 
Function rdepth in R package ``mtfDepth” was used to calculate the RD of each candidate hyperplane. The performance of three algorithms for $\bs{\beta}^*_{RD}$, 
 $\bs{\beta}^*_{PRD}$, 
 and ltsReg,
respectively, is demonstrated in the table \ref{table-comp-time}.  \vs

We generate $1000$ samples $\mb{Z}^{(n)}=\{(\mb{x}'_i, y_i)', i=1,\cdots, n, \mb{x}_i \in \R^{p-1}\}$ from the Gaussian distribution with zero mean vector and $1$ to $p$ as its diagonal entries of the diagonal covariance matrix  for various $n$ and $p$. They are contaminated by $5\%$ i.i.d. normal $p$-dimensional points with mean vector $(10, \cdots, 10)'$ and diagonal covariance matrix with $0.1$ as its diagonal entries. Thus, we no longer have a symmetric errors and homoscedastic variance model
(skewness and heteroscedasticity are allowed for the RD of RH99).\vs

For a general estimator $\mb{T}$, if it 
 is regression equivariant, then we can assume (w.l.o.g.) that the true parameter $\bs{\beta}_0=\mb{0}\in \R^p$.  We calculate
$\mbox{EMSE}:=\frac{1}{R}\sum_{i=1}^R \|\mb{T}_i - \bs{\beta}_0\|^2$, the empirical mean squared error (EMSE) for $\mb{T}$, where
$R = 1000$, $\bs{\beta}_0 = (0, \cdots, 0)'\in \R^{p}$,
 and $\mb{T}_i$ is the realization of $\mb{T}$ obtained from the ith sample with size $n$. The EMSE and the average computation time (in seconds) per sample by different estimators are listed in Table \ref{table-comp-time}.\vs

\begin{table}[h!]
\centering
~~ Table entries: (empirical mean squared error, average time per sample (seconds))
\bec
\begin{tabular}{c c c c c c}
n& method & $p=2$~~~~ & $p=3$~~~~ &$p=4$~~~~& $p=6$~~~~  \\
\hline\\[0.ex]
$40$ 
 & $\bs{\beta}^*_{PRD} $& (0.232, 0.060) &(0.468, 0.261)&(0.723, 0.304)&(1.429, 0.354)\\[.5ex]
 &$\bs{\beta}^*_{RD}$ &(0.243, 0.038)&(0.492, 0.124)& (2.7e+04, 6.542)& (1.717, 9.619)\\[.5ex]
 &ltsReg &(0.380, 0.007)&(0.579,  0.011)& (0.781, 0.010)& (1.434, 0.018)\\[.5ex]\\
$60$ 
 & $\bs{\beta}^*_{PRD} $& (0.160, 0.080) &(0.323, 0.310)&(0.510, 0.445)&(0.894, 0.532)\\[.5ex]
  &$\bs{\beta}^*_{RD}$ &0.172, 0.043)&(0.366, 0.286)& (2565.1, 23.14)& (1.206, 11.82)\\[.5ex]
  &ltsReg &(0.326, 0.007)&(0.475, 0.013)& (0.599, 0.015)& (0.894, 0.024)\\[.5ex]\\
$80$ 
  & $\bs{\beta}^*_{PRD} $& (0.124, 0.100) &(0.260, 0.436)&(0.413, 0.613)&(0.691, 0.634)\\[.5ex]
 &$\bs{\beta}^*_{RD}$&(0.130, 0.047) &(0.291, 0.569)&(2012.6, 58.42)& (1.111, 14.08)\\[.5ex]
 &ltsReg&(0.290, 0.009) &(0.416, 0.018)&(0.506, 0.020)& (0.703, 0.029)\\[.5ex]\\
$100$ 
 & $\bs{\beta}^*_{PRD} $& (0.100, 0.123) &(0.221, 0.528)&(0.346, 0.687)&(0.555, 0.763)\\[.5ex]
  &$\bs{\beta}^*_{RD}$ &(0.109, 0.048)&(0.252, 0.950)& (5.5e+06, 101.8)& (0.963, 16.37)\\[.5ex]
  &ltsReg &(0.252, 0.010)&(0.418, 0.021)& (0.455, 0.024)& (0.578, 0.035)\\[1ex]
\hline
\end{tabular}
\enc
\caption{
Performance of different regression methods for various $n$ and $p$.}
\label{table-comp-time}
\end{table}
\vspace*{-0mm}

\noindent
\tb{Remarks 3.3} Table \ref{table-comp-time} reveals that
\bi
\item[(I)] In terms of the average time consumed per sample, or computation speed, (i) the ltsReg  is the fastest in all cases whereas the $\bs{\beta}^*_{RD}$ is the second fast method when $p$ is $2$, or $3$ (and $n\leq 60$).
  (ii) the  $\bs{\beta}^*_{PRD}$ is the slowest in the $p=2$ and $p=3$ and $(n\leq60)$ cases whereas the $\bs{\beta}^*_{RD}$ unexpectedly becomes the slowest when $(p\geq 4)$. (iii) the $\bs{\beta}^*_{PRD}$ is
  at least 20 (20 to 148) times faster than the $\bs{\beta}^*_{RD}$ when $p>3$.\vs 
  Note the comparison here is somewhat unfair 
  since the ltsReg uses Fortran and
    the $\bs{\beta}^*_{RD}$ or the $\bs{\beta}^*_{PRD}$ employs Rcpp for the background computation. This example also confirms that old Fortran is still an
     excellent programming language for scientific computation.

\item[(II)] Computation speed is just one  important performance criterion. Accuracy or efficiency is another, if not a more important,  one.
In terms of EMSE, there is  an across-the-board winner. That is, the $\bs{\beta}^*_{PRD}$ has the smallest EMSE in all cases
considered (in other words, it has the highest empirical relative efficiency).
\item[(III)] 
The $\bs{\beta}^*_{PRD}$ runs in less than one second in all cases considered. 
     \hfill \pend
\ei

\vs
Overall, Table \ref{table-comp-time} suggests that 
the $\bs{\beta}^*_{PRD}$ is a promising alternative among the competitors in regression.
\vs
All R code (downloadable via the link provided at the end of Example 2.4.1) ran on a desktop Intel(R)Core(TM) i7-2600 CPU @ 3.40GHz.

\vs  The ltsReg has a fairly good finite sample relative efficiency, but it is also notorious
    for its  inefficiency in the asymptotic sense (with an asymptotic efficiency of just $7\%$ (see Stromberg, et al. (2000)). The ltsReg benefits heavily in terms of speed from Fortran’s compilation.

\bec
\begin{figure}[t]
\vspace*{-15mm}
\includegraphics[width=\textwidth]{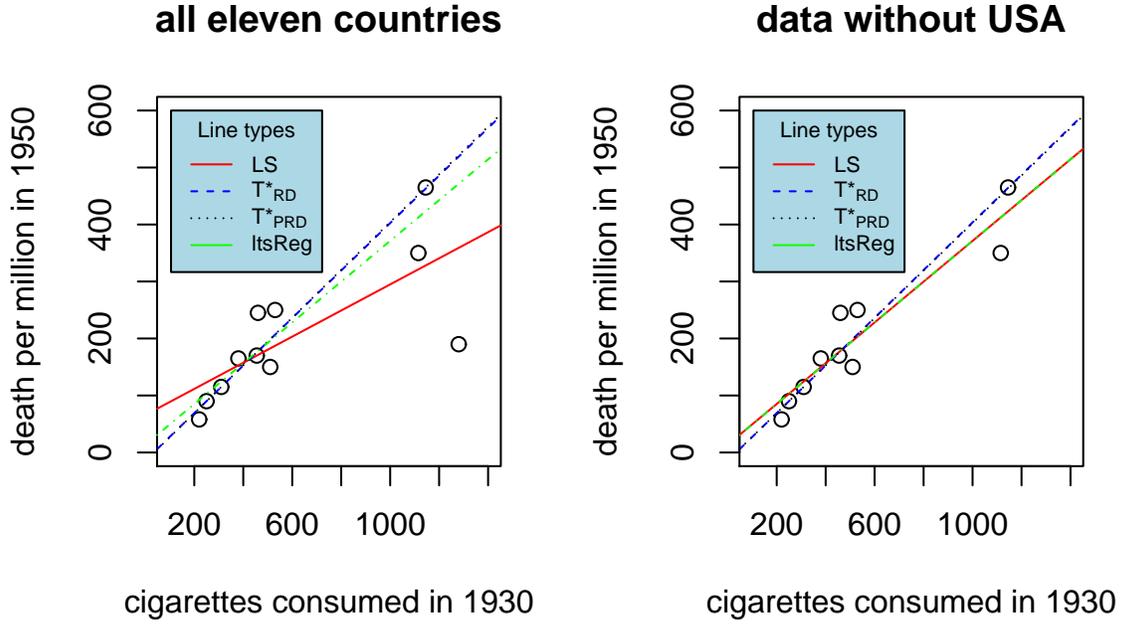}
\vspace*{-10mm}
 \caption{
 {Four regression lines for a dataset with a single outlier (Solid red for the LS, dashed blue for the $\mb{T}^*_{RD}$, dotted black for the $\mb{T}^*_{PRD}$ and dotdash green for the ltsReg). Left: Original eleven countries, lines from the $\mb{T}^*_{RD}$, the $\mb{T}^*_{PRD}$ and the ltsReg are similar while the LS line is attracted by a single country, the USA. Right: The outlier, USA, is removed from the original data, all four lines are very similar and catch the overall pattern.}}
 \label{fig:example-3-1-added}
 \vspace*{-3mm}
\end{figure}
\enc
\vspace*{-2mm}

\bec
\begin{figure}[t]
\vspace*{-15mm}
\includegraphics[width=\textwidth]{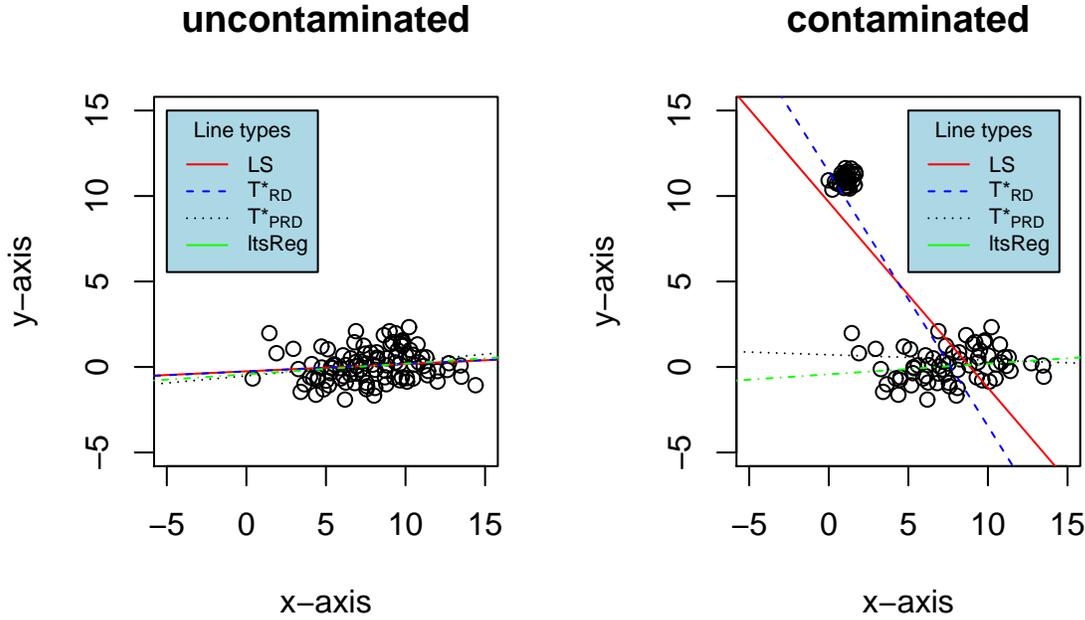}
\vspace*{-10mm}
 \caption{
 {Four regression lines for data with or without contamination (Solid red for the LS, dashed blue for the $\mb{T}^*_{RD}$, dotted black for the $\mb{T}^*_{PRD}$, and dotdash green for the ltsReg). Left: Original 100 normal points, lines from the LS, the $\mb{T}^*_{RD}$, the $\mb{T}^*_{PRD}$ and the ltsReg are similar and catch the overall linear pattern . Right: $34\%$ contaminated data set, both the LS and the $\mb{T}^*_{RD}$ ``break down" while the $\mb{T}^*_{PRD}$ and the ltsReg resist the contamination and still track the major pattern.}}
 \label{fig:example-3-1}
 \vspace*{-3mm}
\end{figure}
\enc
\vspace*{-12mm}

\vs

\subsection{Examples}

\noindent
\tb{Example 3.3.1 Deepest regression line $\mb{T}^*_{PRD}$ for a real data set}
\vs

Lung Cancer and Smoking Data set is composed of per capita consumption of cigarettes in eleven countries in 1930 and the death rates
(number of deaths per million people) from lung cancer in 1950 (see Table 3-3 of Tufte (1974), source: Doll (1955)).\vs

To find out the relationship between death rate and the cigarettes consumed, we first regress the data with the deepest line $\mb{T}^*_{PRD}$, for benchmark and comparison purposes, lines from the $\mb{T}^*_{RD}$ (another line induced from depth), the LS (classical one) and the  lstReg
are also given. In terms of (intercept, slope) form, they are (65.7488570,  0.2291153), (-14.1666667,  0.4166667), (-14.9401198, 0.4191617 ), and (13.553435,  0.357668) for the LS, the $\mb{T}^*_{RD}$, the $\mb{T}^*_{PRD}$ and the ltsReg, respectively. The two depth lines are almost identical whereas the LS line is attracted by a single country USA downward, see the left panel of Figure \ref{fig:example-3-1-added}. \vs

Next we remove the single outlier and repeat the steps above, we have this time (13.553435,  0.357668), (-14.6184633,  0.4182743), (-14.1666667, 0.4166667) and (13.553435  0.357668) for the LS, the $\mb{T}^*_{RD}$, the $\mb{T}^*_{PRD}$ and the ltsReg, respectively. The ltsReg and the LS are the same as the previous lstReg with USA included and the $\mb{T}^*_{RD}$ and the $\mb{T}^*_{PRD}$ are almost the same with the $\mb{T}^*_{PRD}$ being exactly the same as the previous $\mb{T}^*_{RD}$ with USA included, see the right panel of Figure \ref{fig:example-3-1-added}.\vs

Overall, this example indicates that a single outlier can drastically affect the LS line and distinguishes the LS line from other robust lines, but it can not differentiate the other three. \hfill \pend
\vs
\noin
\noindent
\tb{Example 3.3.2 Performance of the deepest line of  $\mb{T}^*_{PRD}$ versus  $\mb{T}^*_{RD}$}
\vs
\noin
Here we first generate $100$ points $\mb{Z}_i=(x,y)'$ from the bivariate normal distribution $\mb{N}(\bs{\mu}, \Sigma)$, where
\[
\bs{\mu}=\begin{pmatrix}8\\0\end{pmatrix},~~~~\Sigma=\begin{pmatrix}9&0.9\\0.9 &1\end{pmatrix}. \]
\vs
\noin
Among the 100 points, we randomly select  $34$ points and replace them by 34 other points from another bivariate normal distribution $\mb{N}(\bs{\mu}_c, \Sigma_c)$ with
\[\bs{\mu}_c=\begin{pmatrix}1\\11\end{pmatrix},~~~~ \Sigma_c=\begin{pmatrix}0.1&0\\0&0.1\end{pmatrix},
\]
Thus, we have a $34\%$ replacement-contamination data set.\vs

First: w.r.t. the un-contaminated data set, we compute the deepest regression line induced from the PRD and then the competitor line induced from the RD of RH99. 
Again we also calculated the LS and the ltsReg lines.
The four lines in (intercept, slope) form are (-0.2533241,  0.0431503), (-0.25902968, 0.04446287), (-0.51728622,  0.08431267) and (-0.42734074,  0.06248309) for the LS, the $\mb{T}^*_{RD}$, the $\mb{T}^*_{PRD}$ and  the ltsReg, respectively. They are almost identical as shown in the left panel of Figure \ref{fig:example-3-1}. All four seem to be useful, catching the overall linear pattern.\vs

Second: w.r.t. the replacement-contaminated data set, we also compute the four lines. They are (9.643466, -1.085616), (11.407796, -1.487017), (0.71612780, -0.03124584) and (-0.42734074, 0.06248309) for the LS, the $\mb{T}^*_{RD}$, the $\mb{T}^*_{PRD}$ and the ltsReg, respectively.
They differ in obvious manners as shown in the right panel of Figure (\ref{fig:example-3-1}). Both the LS and the $\mb{T}^*_{RD}$ lines  break down (attracted by the cloud of contamination) whereas the $\mb{T}^*_{PRD}$ and the ltsRg can resist the $34\%$ contamination (in fact up to $50\%$) and catch the major pattern and continue to provide a useful regression line.
\vs
The computations in the example above (and below) are  carried out with the R programming language for two reasons: (i) available code (package: mrfDepth) for the RD of RH99 is in R  and (ii) fair comparisons. R code is available (see the link posted at the end of Example 2.4.1).
\hfill \pend
\vs

\noin
\tb{Remarks 3.2}:
 \begin{itemize}
\item[(I)] Example 3.3.2 confirms the theoretical results in Z18b. That is, the deepest regression line or hyperplane induced from the RPD is a robust alternative to the traditional LS lines or hyperplanes and  has a higher asymptotical breakdown point (ABP) ($50\%$) than the leading depth median ($33\%$), the deepest regression estimator induced from the RD of RH99. Note that with an appropriate trimming rate the least trimmed squares line possesses the best possible ABP whereas if the rate tends to $0\%$ it leads to the LS line or hyperplane having $0\%$ ABP since just one outlier can ruin them.
\vs
\item[(II)] Robustness does not work in tandem with efficiency. So the key question is: Are the deepest projection regression lines or hyperplanes ($\mb{T}^*_n$) efficient? In the following, ltsReg is excluded for a pure apples vs apples comparison (depth median vs depth median)
    and since it is notorious
    for its inefficiency (with asymptotic efficiency of just $7\%$ (see Stromberg, et al.(2000)).
    \hfill \pend
\end{itemize}

\vs

\begin{table}[h!]
\centering
Replication $1000$ times,  $n=65$  \\[1ex]
\bec
\begin{tabular}{c c c c }
 Performance criteria~~~~~~ & $\bs{\beta}^*_{PRD}$~~~~&$\bs{\beta}^*_{RD}$ 
  \\[.5ex]
\hline\\[0.ex]
&{\bf{Case I}}& $p=3$ &  \\[1.5ex]
EMSE &0.09328212 &0.11425414 
\\[.5ex]
Time consumed per sample  &0.31453851&0.33221344 
\\[.5ex]
\hline\\[.ex]
&{\bf{Case II}}& $p=4$ & \\[1.5ex]
EMSE&0.1516812&3265582 
\\[.5ex]
Time consumed per sample &0.25505812 &25.63505434 
\\[.5ex]
\hline\\[.ex]
&{\bf{Case III}}& $p=5$ & \\[1.5ex]
EMSE&0.2302744 &0.2706360
\\[1.5ex]
Time consumed per sample &0.21889922 &6.42221254 
\\[.5ex]
\hline\\[0.ex]
\end{tabular}
\enc
\vspace*{-9mm}
\caption{
Performance of different regression depth medians for the three true $\bs{\beta}_0$s.}
\label{table-3-betas}
\vspace*{-0mm}
\end{table}

\noin
\tb{Example 3.3.3} Now we investigate the performance of the two regression depth medians ($\bs{\beta}^*_{PRD}$, and $\bs{\beta}^*_{RD}$) in a slightly different setting. We generate $1000$ samples $\{(\mb{x}'_i, y_i)' \in \R^p\}$ with a fixed sample size $65$ from an assumed model: $y=\bs{\beta_0}'\mb{x}     
+e$, where $\mb{x}=(1,x_1,\cdots, x_{p-1})'$ and $\bs{\beta_0}=(\beta_0,\cdots, \beta_{p-1})'$ are in $\R^p$ and $x_i$ and e are from either a Cauchy or a standard Gaussian distribution.\vs  We list the average time consumed (in seconds) per sample and the EMSE (the same formula as before) for the two methods
with respect to different $\bs{\beta_0}$'s in Table \ref{table-3-betas}.  {\bf{Case I}} $\bs{\beta_0}=(-2, 0.1,1)'$, all $x_i$ and $e$ are from $N(0,1)$ distribution.
{\bf{Case II}} $\bs{\beta_0}=(-2, 0.1,1, 5)'$, $x_1$ is from $N(0,1)$ and all other $x_i$ and $e$ are from the Cauchy distribution. {\bf{Case III}} $\bs{\beta_0}=(50, 0.1, -2, 15, 100)'$, all $x_i$ and $e$ are from $N(0,1)$.
\vs
Inspecting table \ref{table-3-betas} reveals that (i) the $\bs{\beta}^*_{PRD}$ is 
 faster than the $\bs{\beta}^*_{RD}$ in all cases; it is $100.5$ and $29.34$ times faster in cases $p=4$ and $p=5$, respectively. (ii) the $\bs{\beta}^*_{PRD}$ has a smaller EMSE in all cases. (iii) The sample variance (or more precisely EMSE) of the PRD induced median increases when $p$ increases whereas the time consumed per sample for the fixed sample size by the $\bs{\beta}^*_{PRD}$ decreases.\vs

\begin{figure}[ht]
    \centering
    \begin{subfigure}[ht]{0.3\textwidth}
        \includegraphics[width=4cm, height=4cm]{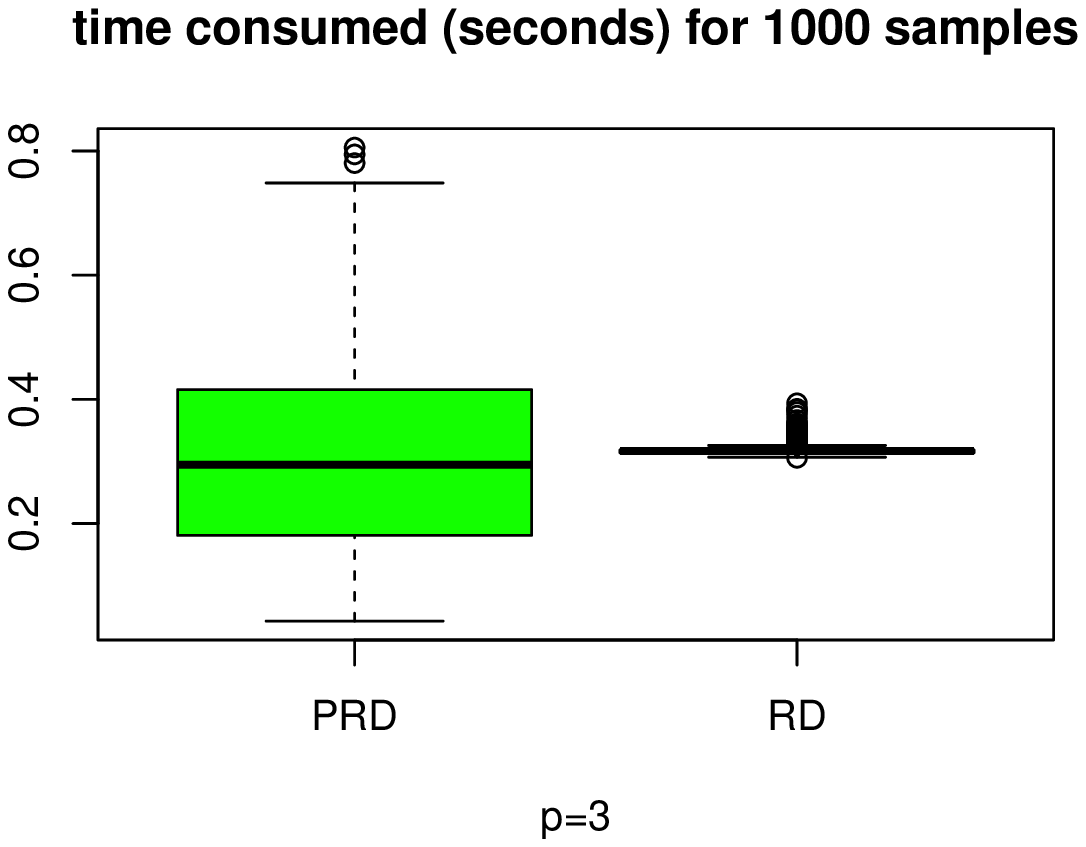}
        \caption{p=3}
        \label{fig:no-camtami}
    \end{subfigure}
     \begin{subfigure}[ht]{0.3\textwidth}
        \includegraphics[width=4cm, height=4cm]{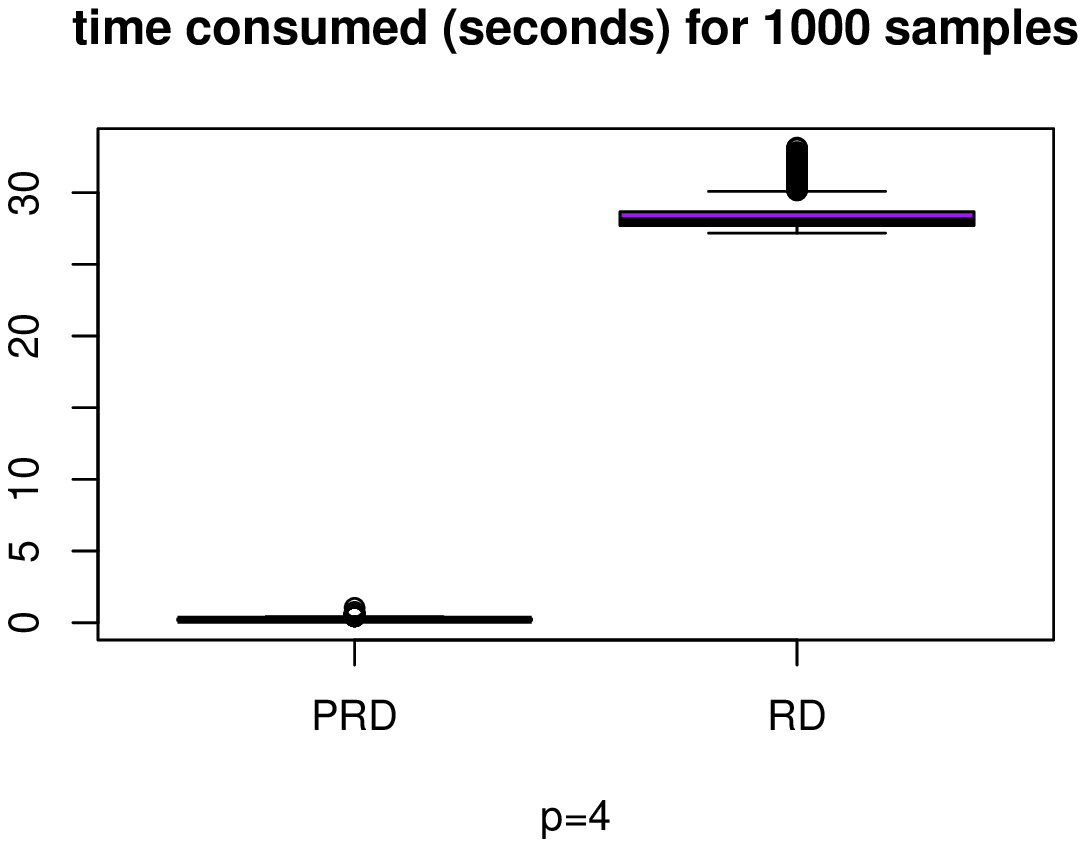}
        \caption{p=4}
        \label{fig:one-contami}
    \end{subfigure}
    \begin{subfigure}[ht]{0.3\textwidth}
        \includegraphics[width=4cm, height=4cm]{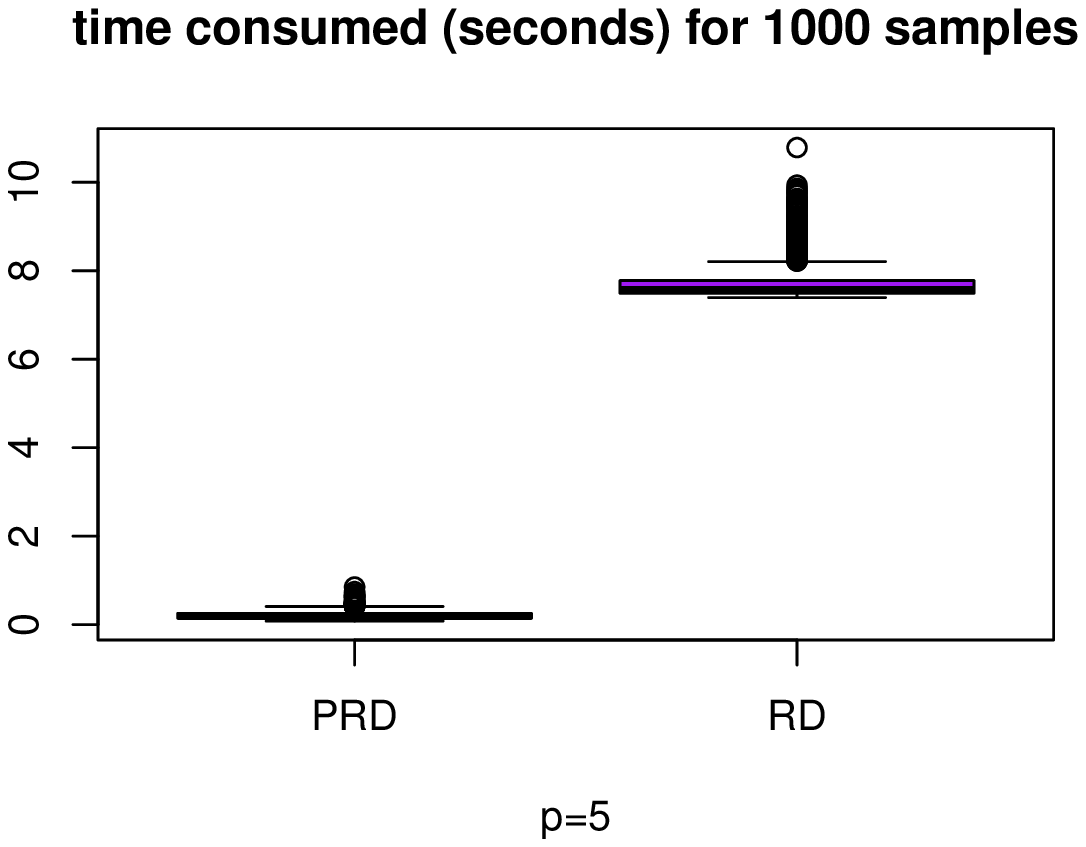}
        \caption{p=5}
        \label{fig:contami}
    \end{subfigure}
    \caption{
    Time consumed per sample by two depth induced deepest regression estimators $\bs{\beta}^*_{PRD}$ (green box) and $\bs{\beta}^*_{RD}$ (purple box) for the three $\bs{\beta}_0$ cases.}
\label{fig-boxplot}
\end{figure}

\begin{figure}[ht]
    \centering
    \begin{subfigure}[ht]{0.3\textwidth}
        \includegraphics[width=4cm, height=4cm]{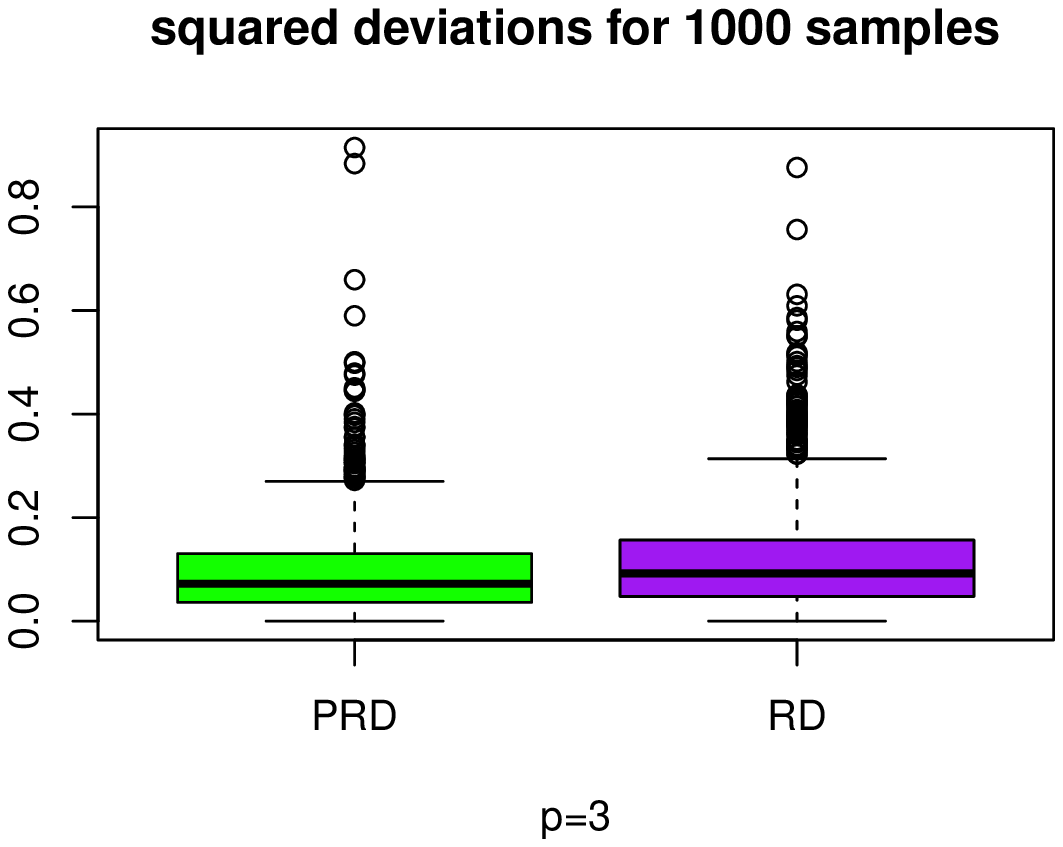}
        \caption{p=3}
        \label{fig:no-camtami}
    \end{subfigure}
     \begin{subfigure}[ht]{0.3\textwidth}
        \includegraphics[width=4cm, height=4cm]{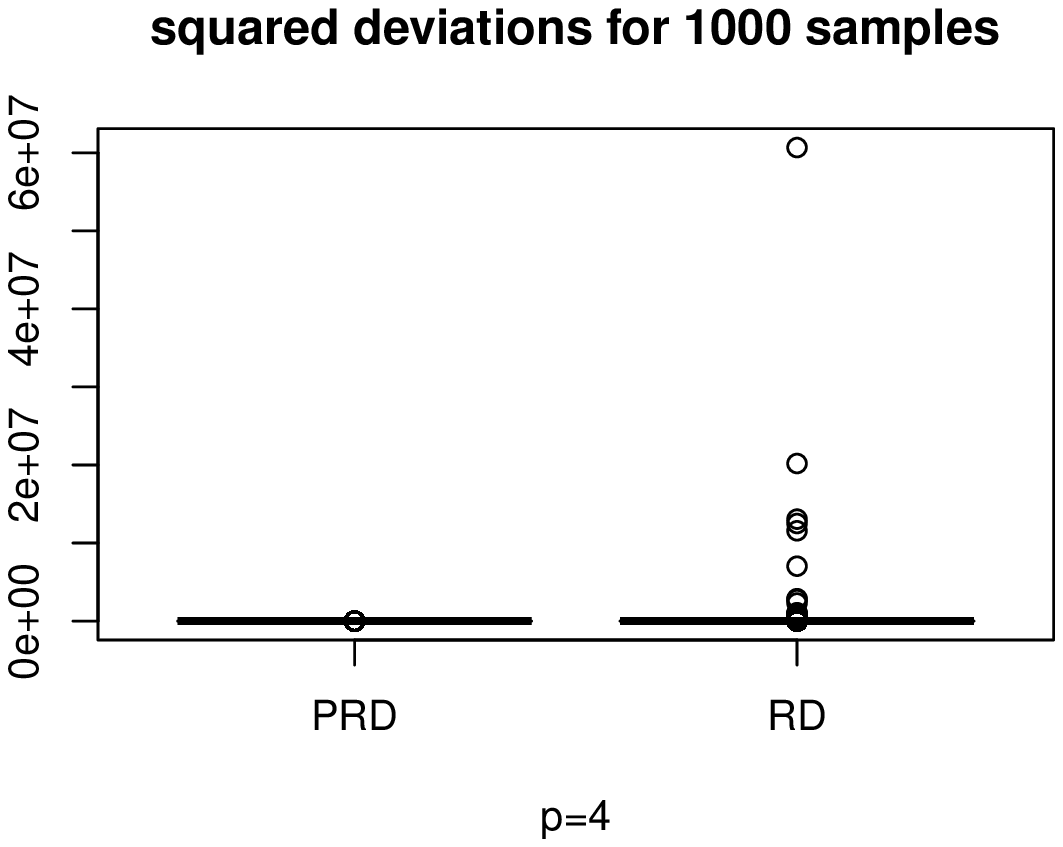}
        \caption{p=4}
        \label{fig:one-contami}
    \end{subfigure}
    \begin{subfigure}[ht]{0.3\textwidth}
        \includegraphics[width=4cm, height=4cm]{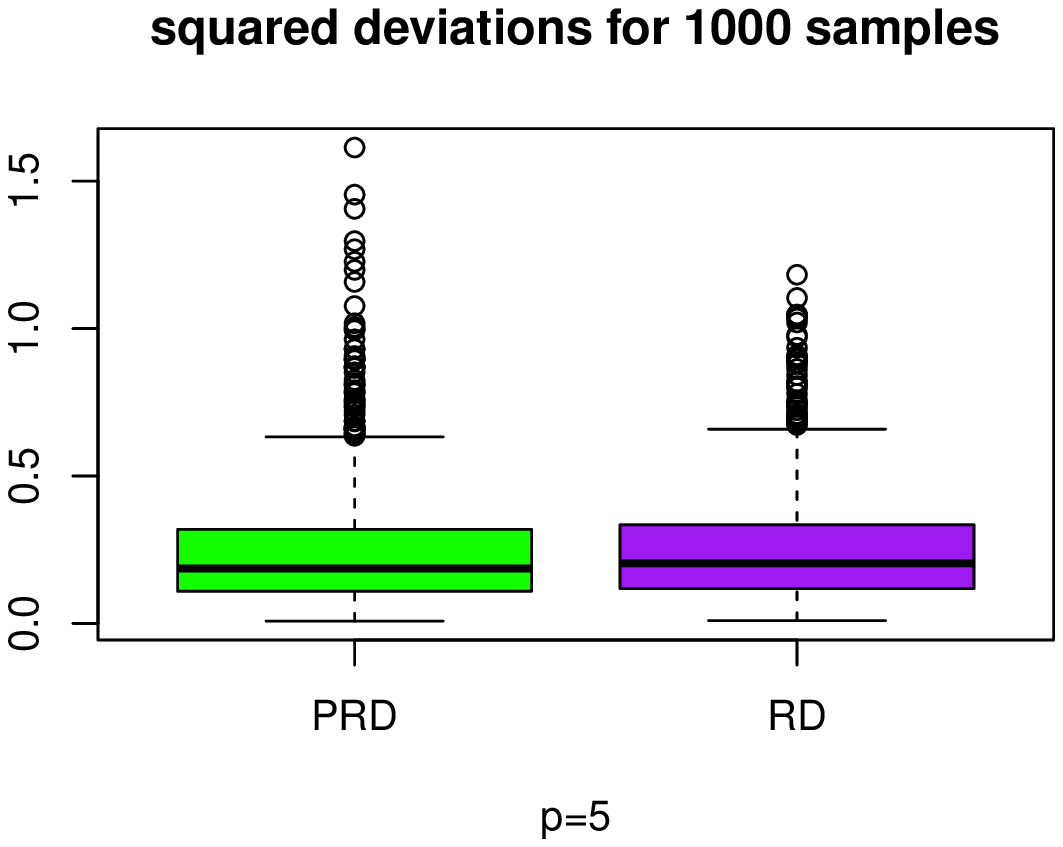}
        \caption{p=5}
        \label{fig:contami}
    \end{subfigure}
    \caption{Squared-deviations $\|\bs{\beta}^*_i-\beta_0\|^2$ per sample by two depth induced deepest regression estimators $\bs{\beta}^*_{PRD}$ (green box) and $\bs{\beta}^*_{RD}$ (purple box) for the three $\bs{\beta}_0$ cases.}
\label{fig-squared-deviations}
\end{figure}

Numerical summary results in table \ref{table-3-betas} for 1000 samples are also displayed graphically in terms of their distributions in Figures \ref{fig-boxplot} and \ref{fig-squared-deviations}. 
The times consumed by two methods for each of 1000 samples are displayed in Figure \ref{fig-boxplot} with boxplots. Inspecting the Figure immediately reveals that in terms of time consumed per sample, the $\bs{\beta}^*_{PRD}$
is faster than the $\bs{\beta}^*_{RD}$ in all, especially $p=4$ and $p=5$ cases. \vs

\vs
\bec
\vspace*{-7mm}
\enc
The distributions of the squared deviations $\|\bs{\beta}^*_i-\beta_0\|^2$ per sample for two methods  $\bs{\beta}^*_{PRD}$ and $\bs{\beta}^*_{RD}$ are displayed in Figure \ref{fig-squared-deviations}. The figure clearly indicates that in the cases $p=3$ and $p=5$,
the median of the squared-deviations of the $\bs{\beta}^*_{PRD}$ is the smaller one whereas it is also true in the case $p=4$ but less clear. Notice that in the latter case, there is an obvious outlier for the $\bs{\beta}^*_{RD}$; it is greater than $6.0\times 10^7$, which explains why the mean of
the squared-deviations (empirical mean squared error, EMSE) in table \ref{table-3-betas} for the $\bs{\beta}^*_{RD}$ is huge. \hfill \pend
\vs
\bec
\vspace*{-10mm}
\enc
\vs
All results above and below are obtained on a desktop Intel(R)Core(TM) i7-2600 CPU @ 3.40GHz. 
R code in this and the next few sections is downloadable via the link listed at the end of Example 2.4.1.\vs

\section{Three estimators induced from PRD}
From table \ref{table-comp-time} we see that the $\bs{\beta}^*_{PRD}$ is slower than $\bs{\beta}^*_{RD}$ in the case $p=2$ or $p=3 (n\leq 60)$.
 Are there any PRD induced estimators that run even faster than the $\bs{\beta}^*_{PRD}$?
 By reviewing the steps in section 3.1 for the computation of $\bs{\beta}^*_{PRD}$,
the answer is yes. There are obviously other projection regression depth (PRD) induced estimators that can be computed even faster.
\vs

\vspace*{-0mm}
\begin{table}[h!]
\centering
~~ Table entries: (empirical mean squared error, average time per sample (seconds))
\bec
\begin{tabular}{c c c c c c}
n& method & $p=2$~~~~ & $p=3$~~~~ &$p=4$~~~~& $p=6$~~~~  \\
\hline\\[0.ex]
$40$ 
 & $\bs{\beta}^*_{PRD} $& (0.237, 0.062) &(0.448, 0.142)&(0.736, 0.208)&(1.373, 0.343)\\[.5ex]
 & $\bs{\beta}^*_{PRD1} $& (0.244, 0.023) &(0.481, 0.040)&(0.831, 0.068)&(1.646, 0.142)\\[.5ex]
 & $\bs{\beta}^*_{PRD2} $& (0.268, 0.023) &(0.489, 0.040)&(0.882, 0.068)&(1.431, 0.142)\\[.5ex]
 & $\bs{\beta}^*_{PRD3} $& (0.258, 0.023) &(0.476, 0.040)&(0.771, 0.068)&(1.375, 0.142)\\[.5ex]
 &$\bs{\beta}^*_{RD}$ &(0.240, 0.040)&(0.466, 0.124)& (3195.3, 6.507)& (1.678, 9.382)\\[.5ex]\\
$60$ 
 & $\bs{\beta}^*_{PRD} $& (0.157, 0.082) &(0.329, 0.187)&(0.519, 0.268)&(0.923, 0.193)\\[.5ex]
 & $\bs{\beta}^*_{PRD1} $& (0.167, 0.031) &(0.363, 0.051)&(0.613, 0.090)&(1.139, 0.088)\\[.5ex]
 & $\bs{\beta}^*_{PRD2} $& (0.188, 0.031) &(0.484, 0.051)&(0.603, 0.090)&(1.131, 0.088)\\[.5ex]
 & $\bs{\beta}^*_{PRD3} $& (0.175, 0.031) &(0.446, 0.051)&(0.568, 0.090)&(1.075, 0.088)\\[.5ex]
  &$\bs{\beta}^*_{RD}$ &(0.165, 0.043)&(0.350, 0.300)& (4703.0, 21.18)& (1.337, 8.585)\\[.5ex]\\
$80$ 
  & $\bs{\beta}^*_{PRD} $& (0.128, 0.101) &(0.261, 0.441)&(0.412, 0.611)&(0.666, 0.288)\\[.5ex]
  & $\bs{\beta}^*_{PRD1} $& (0.134, 0.040) &(0.297, 0.095)&(0.492, 0.165)&(0.832, 0.129)\\[.5ex]
  & $\bs{\beta}^*_{PRD2} $& (0.165, 0.040) &(0.315, 0.095)&(0.509, 0.165)&(0.872, 0.129)\\[.5ex]
   & $\bs{\beta}^*_{PRD3} $& (0.147, 0.040) &(0.302, 0.095)&(0.481, 0.165)&(0.830, 0.129)\\[.5ex]
 &$\bs{\beta}^*_{RD}$&(0.132, 0.047) &(0.291, 0.583)&(4446.2, 58.50)& (1.050, 10.64)\\[.5ex]\\
$100$ 
 & $\bs{\beta}^*_{PRD}$& (0.109, 0.121) &(0.218, 0.301)&(0.361, 0.719)&(0.551, 0.338)\\[.5ex]
 & $\bs{\beta}^*_{PRD1} $& (0.117, 0.048) &(0.247, 0.086)&(0.439, 0.202)&(0.682, 0.148)\\[.5ex]
 & $\bs{\beta}^*_{PRD2}$& (0.153, 0.048) &(0.275, 0.086)&(0.467, 0.202)&(0.851, 0.148)\\[.5ex]
 & $\bs{\beta}^*_{PRD3}$& (0.142, 0.048) &(0.263, 0.086)&(0.437, 0.202)&(0.771, 0.148)\\[.5ex]
  &$\bs{\beta}^*_{RD}$ &(0.115, 0.050)&(0.240, 0.960)& (2427164, 113.4)& (0.970, 12.24)\\[.5ex]
\hline
\end{tabular}
\enc
\caption{Performance of regression depth induced estimators for various $n$ and $p$.}
\label{table-comp-time-2}
\end{table}
\vspace*{-3mm}
\vs
 The first one adds no extra computation cost to the already obtained candidate $\bs{\beta}$ matrix $S_{\bs{\beta}}$ and their UFs, it is just the deepest $\bs{\beta}$ with the minimum UF in the matrix $S_{\bs{\beta}}$, denote it  as $\bs{\beta}^*_{PRD1}$. The second one is the plain average of the deepest $(p+1)$ $\bs{\beta}$s from the $S_{\bs{\beta}}$, denote it  as $\bs{\beta}^*_{PRD2}$.
 The third one is an UF weighted estimator defined below, denote it  as $\bs{\beta}^*_{PRD3}$,
\be
\bs{\beta}^*_{PRD3}=\frac{\sum_{i=1}^{(p+1)}w(\rho_i)\bs{\beta}_{(i)}}{\sum_{i=1}^{(p+1)}w(\rho_i)}, \label{wprd.eqn}
\ee
where $\rho_i=\mbox{UF}(\bs{\beta}_{(i)})$ and $\bs{\beta}_{(1)},\cdots,\bs{\beta}_{(p+1)}$ are the first $(p+1)$ deepest $\bs{\beta}$s (with the least UF) in the $S_{\bs{\beta}}$ and the weight function $w$ is defined as follows:
\be
w(r)={\bf{I}}(r\leq r_0)+{\bf{I}}(r>r_0)\frac{exp~\Big(k\big(2r_0/r-(r_0/r)^2\big)\Big)-1}{exp~ (k)-1}, \label{weight.eqn}
\ee
with the tuning parameters 
$k=3$ and $r_0=\rho_{(p-1)}$, the $(p-1)$th smallest UF among the $(p+1)$ minimum UFs.
For more discussions on this weight function and the tuning parameters, refer to Zuo (2003) and Z19b.
\vs

These estimators obviously can run faster than $\bs{\beta}^*_{PRD}$ since they skip the time-consuming step of searching over the convex hull. One naturally wonders what are their EMSE's?
We investigate the performance of 
$\bs{\beta}^*_{PRD}$, $\bs{\beta}^*_{PRD1}$, $\bs{\beta}^*_{PRD2}$, and $\bs{\beta}^*_{PRD3}$ which is reported in table \ref{table-comp-time-2}. For the benchmark purpose, the depth median: $\bs{\beta}^*_{RD}$ of RH99 is included in the comparison. $1000$ samples are generated with the same scheme as that for table  \ref{table-comp-time}. \vs

Inspecting table \ref{table-comp-time-2} immediately reveals that (i) the $\bs{\beta}^*_{PRD}$  has the smallest EMSE in  all cases
and it can be 
 faster  than  the $\bs{\beta}^*_{RD}$ which is the slowest in $p=3 (n>40)$, $p=4$, $p=6$  cases; (ii) the  $\bs{\beta}^*_{PRD1}$, the $\bs{\beta}^*_{PRD2}$ and the $\bs{\beta}^*_{PRD3} $ are the fastest 
and they are currently regarded as having the same speed (all dependent on the given matrix $S_{\bs{\beta}}$ of candidate $\bs{\beta}$s and their unfitness and then on the sorted values of their unfitness);
(iii)
 among the three, the deepest of all $\bs{\beta}$s in $S_{\bs{\beta}}$, $\bs{\beta}^*_{PRD1}$, and the depth weighted deepest $(p+1)$ $\bs{\beta}$s, $\bs{\beta}^*_{PRD3}$,  seem to perform better than the plain average, $\bs{\beta}^*_{PRD2}$, which seems to perform the worst in most cases.
Furthermore, our empirical evidence indicates that $\bs{\beta}^*_{PRD3}$ performs even better when $p$ increases (say $p\geq 8$).   (vi) Overall, $\bs{\beta}^*_{PRD}$ should be recommended among the five depth induced regression estimators; it becomes empirically the same as $\bs{\beta}^*_{PRD1}$ for large $p$ (e.g. $p=20$, $n=40, 60, 80$); the second most impressive one is the
$\bs{\beta}^*_{PRD3}$, and  $\bs{\beta}^*_{PRD2}$ seems to be mediocre.
\vs

\section{Finite sample relative efficiency of deepest projection regression lines/hyperplanes}

Example 3.3.2 confirms that the $\mb{T}^*_{PRD}$ (or $\mb{T}^*_n$ in the empirical case) has a higher breakdown point than that of the leading regression depth induced median,  the $\mb{T}^*_{RD}$.  Robustness, however, is just one performance criterion for an estimator. Efficiency, if not more important, is another major performance measure.
 One naturally wonders whether the $\mb{T}^*_{PRD}$ is inferior to the $\mb{T}^*_{RD}$ w.r.t. the efficiency criterion.\vs
The immediate answer is no based on the evidence demonstrated in tables 5, 6 and 7 since the $\mb{T}^*_{PRD}$ has a smaller EMSE than that of the $\mb{T}^*_{RD}$ uniformly over all cases considered. To confirm this empirical observation in $p=2$ case, we now carry out a small scale simulation study.\vs
\begin{table}[h!]
\centering 
{\tb ($\epsilon=0$) }\\
 Empirical mean squared error and relative efficiency of the $\mb{T}^*_{RD}$ and the $\mb{T}^*_{PRD}$ w.r.t. the LS estimator   \\[1ex]
\begin{tabular}{ccccc}
\hline\\[.0ex]
n &measures & $\mb{T}^*_{RD}$& $\mb{T}^*_{PRD}$   & LS\\[2ex]
10& EMSE &0.5987264&0.3723071&0.2653862\\[1ex]
  &   RE & $44\%$ & $71\%$  & $100\%$ \\[2ex]
20& EMSE& 0.2358544 &0.1571197&0.1104146\\[1ex]
  &   RE & $47\%$ & $70\%$  & $100\%$ \\[2ex]
40&EMSE&0.10163933&0.07492950&0.05287073\\[1ex]
&   RE & $52\%$ & $71\%$ & $100\%$  \\[2ex]
80&EMSE&0.04893200&0.04060671& 0.02597673\\[1ex]
&   RE &$53\%$ &$64\%$ & $100\%$ \\[2ex]
100&EMSE &0.03196556&0.02535978&0.01679633\\[1ex]
&   RE & $53\%$ & $66\%$ & $100\%$  \\[1ex]
\hline
\end{tabular}
\caption{
Relative efficiency of the $\mb{T}^*_{RD}$ and the $\mb{T}^*_{PRD}$ for a normal model with $0\%$ contamination.}
\label{table-3}
\vspace*{-2mm}
\end{table}

\begin{table}[h!]
\vspace*{-0mm}
\centering 
{\tb($\epsilon=0.1$) }\\[1ex]
\begin{tabular}{ccccc}
\hline\\[.0ex]
n &measures & $\mb{T}^*_{RD}$& $\mb{T}^*_{PRD}$  & LS\\[2ex]
10& EMSE &0.6612575&0.6181737& 0.6373658\\[1ex]
  &   RE & $96\%$ & $103\%$ & $100\%$ \\[2ex]
20& EMSE& 0.3396453 &0.3247345&0.5179225\\[1ex]
  &   RE &$152\%$ &$159\%$ & $100\%$ \\[2ex]
40&EMSE&0.1613517&0.1525281&0.4475525\\[1ex]
&   RE &$277\%$ &$293\%$ &  $100\%$  \\[2ex]
80&EMSE&0.10348167&0.09775415&0.43277938\\[1ex]
&   RE &$418\%$ &$442\%$ & $100\%$ \\[2ex]
100&EMSE &0.09702797&0.08947668&0.42298543\\[1ex]
&   RE &$436\%$ & $473\%$ & $100\%$ \\[1ex]
\hline
\end{tabular}
\caption{
Relative efficiency of the $\mb{T}^*_{RD}$ and the $\mb{T}^*_{PRD}$ for a normal model with $10\%$ contamination.}
\label{table-4}
\end{table}

In the following we investigate via a simulation study the finite-sample relative efficiency of the deepest lines $\mb{T}^*_{RD}$ and $\mb{T}^*_{PRD}$ w.r.t. the benchmark, the classical least squares line. The latter is optimal with normal models by the Gauss-Markov theorem. We generate $R=1,000$ samples from the simple linear regression model: $y_i=\beta_0+\beta_1 x_i+e_i, i=1,2,\cdots, n,$
with different sample sizes $n$ (see Tables \ref{table-3} and \ref{table-4}), where $e_i \sim N(0,\sigma^2)$.
\vs

In light of the regression equivariance of the deepest regression estimators (see Z18a), we can assume w.l.o.g. that the true parameter $\bs{\beta}_0=(\beta_0, \beta_1)'=(0,0)'$. We generate $(x_i, y_i)'$ from an $\epsilon\%$ contaminated normal model $(1-\epsilon)N( (0,0)', I_{2\times 2})+\epsilon \delta_{(4,4)'}$
with $\epsilon=0$ (a pure normal model, no contamination) and  $\epsilon=0.1$ (a $10\%$ contaminated normal model), where $\delta_{\mb{Z}}$ is a point mass contaminating distribution at point $\mb{Z} \in \R^2$. \vs

 For a general estimator $\mb{T}$,
the relative efficiency (RE) of the $\mb{T}$ is obtained by dividing the EMSE of the LS estimator by that of the $\mb{T}$.  Tables \ref{table-3} (a pure normal model case) and \ref{table-4} (a normal model with $10\%$ contamination) demonstrate the results with various $n$s.
\vs

Table \ref{table-3} reveals that (i) the $\mb{T}^*_{PRD}$ is uniformly more efficient than the $\mb{T}^*_{RD}$ for all $n$;
(ii) the limited numbers in Table \ref{table-3} give a false impression that the efficiency of the $\mb{T}^*_{RD}$ increases forever as n increases. This is not true since when $n=200$ the efficiency of the $\mb{T}^*_{RD}$ is still just $52\%$; (iii) as expected, the EMSE of any line decreases when $n$ increases.  \vs

On the other hand, with the $10\%$ contaminated normal model, Table \ref{table-4} shows that (i) when $n=10$, there is just one point that is contaminated. The classical least squares line as well as  the line $\mb{T}^*_{RD}$ are drastically affected by just one contaminated point, nevertheless. They are less efficient than the deepest projection regression depth line. It is surprising that the line produced by the $\mb{T}^*_{RD}$ is sensitive to just one point contamination and is even less efficient than the LS line (This phenomenon can be explained by the low \emph{finite sample} breakdown point of the $\mb{T}^*_{RD}$, which could be much lower than the ABP $1/3$, see Zuo (2020));  (ii) when $n$ increases, the efficiency of both deepest depth lines increases and are much higher than that of the LS line; (iii) the $\mb{T}^*_{PRD}$ is more efficient than the $\mb{T}^*_{RD}$  uniformly for all $n$; (iv) the EMSE of any line decreases when $n$ increases; (v) the efficiency of the deepest lines increases as $n$ increases, for example, when $n=200$, the efficiency of the $\mb{T}^*_{RD}$ will be
$525\%$. 

\vs
\section{Concluding remarks}
The maximum projection regression depth estimator is a robust alternative to the classical least squares estimator. It possesses the best asymptotic breakdown point, a bounded influence function, and a very high finite sample replacement breakdown point (see Zuo (2019a)).
\vs

This article addresses the computation issues of the unfitness (UF), or equivalently the projection regression depth (PRD), and  the PRD induced regression median, the maximum projection depth estimator. Exact and approximate algorithms are proposed and investigated.
Compared with the leading regression depth notion  RD and its induced median $\mb{T}^*_{RD}$ (RH99), the $\mb{T}^*_{PRD}$ is more computationally intensive 
$O(Rp N_{\bs{\beta}}(p^2+nN_{\mb{v}}N_{Iter}))$ versus $O(p(pn+N_{Iter}n+n\log n))$ of the $\mb{T}^*_{RD}$ (Van Alest et al (2002)). The $\mb{T}^*_{PRD}$, however, is not only more robust but also more efficient. \vs
The article also introduces three PRD induced estimators that can run very fast. 
 These estimators have a low level of empirical mean squared errors 
 while satisfying regression, scale, and affine equivariance.
 The three estimators are expected to be highly robust, just like the $\bs{\beta}^*_{PRD}$ in Z19b with a high finite sample breakdown point.
 \vs
The advantage (or disadvantage) of the PRD comes from its definition; it is defined based on the magnitude of residuals whereas the RD is defined purely on the sign of residuals. The latter results in a low breakdown point and efficiency, and non-uniqueness of the $\mb{T}^*_{RD}(F^n_{\mb{Z}})$ (the average might not have the maximum depth, see Van Aelst et al (2002) and Mizera and Volauf (2002)) whereas the $\mb{T}^*_{PRD}(F^n_{\mb{Z}})$ is unique (see Zuo (2019b)).
\vs However, the RD also gains some unique features. For example, among others, the RD has the
unique invariance property under the monotone transformations (see RH99); it 
is less difficult to compute and its definition does not require symmetry or homoscedasticity (see RH99).
\vs

\begin{center}
{\textbf{\large Acknowledgments}}
\end{center}
The author thanks Hanwen Zuo, Hanshi Zuo, and Professor Emeritus James Stapleton for their careful proofreading.
He thanks Yan-Han Chen and Dr. Wei Shao for their proofreading and useful discussions on C++, R, and  Rcpp programming. Special thanks go to two anonymous referees for their insightful comments and constructive suggestions, all of which have significantly improved the manuscript.
\end{document}